\PassOptionsToPackage{table,xcdraw}{xcolor}
\documentclass[sigconf,nonacm,balance=false]{acmart}

\usepackage{xcolor}

\usepackage{popets}

\setcopyright{popets}
\copyrightyear{YYYY}

\acmYear{YYYY}
\acmVolume{YYYY}
\acmNumber{X}
\acmDOI{XXXXXXX.XXXXXXX}
\acmISBN{}
\acmConference{Proceedings on Privacy Enhancing Technologies}
\usepackage{longtable}
\usepackage{supertabular}
\usepackage{multicol}
\usepackage{tabularx}
\usepackage{paracol}
\usepackage{wrapfig}
\usepackage{caption}
\usepackage{fontawesome5}
\usepackage{pifont}
\usepackage{pifont} 
\usepackage{graphicx} 
\usepackage{longtable}
\usepackage{float}
\usepackage{booktabs}
\usepackage[table]{xcolor}
\usepackage{makecell} 
\usepackage{colortbl} 
\usepackage{hyperref}
\usepackage{amsmath}
\usepackage{url}
\usepackage{enumitem}
\usepackage[most]{tcolorbox}
\usepackage[framemethod=TikZ]{mdframed} 

\usepackage{array}

\definecolor{darkerGreen}{rgb}{0, 0.8, 0}
\definecolor{red}{rgb}{1, 0, 0} 

\makeatletter
\DeclareRobustCommand{\circbullet}{\mathbin{\vphantom{\circ}\text{\circbullet@}}}
\newcommand{\circbullet@}{%
  \check@mathfonts
  \m@th\ooalign{%
    \clipbox{0 0 0 {\dimexpr\height-\fontdimen22\textfont2}}{$\bullet$}\cr
    $\circ$\cr
  }%
}
\makeatother

\hypersetup{
    colorlinks=true,
    linkcolor=darkerGreen,  
    urlcolor=darkerGreen,   
    citecolor=red           
}

\title{``We Need a Standard'': Toward an Expert–Informed Privacy Label for Differential Privacy}

\author{Onyinye Dibia}
\authornote{Lead author.}
\email{Onyinye.Dibia@uvm.edu}
\affiliation{\institution{University of Vermont}\city{}\state{}\country{}}

\author{Mengyi Lu}
\email{Mengyi.Lu@uvm.edu}
\affiliation{\institution{University of Vermont}\city{}\state{}\country{}}

\author{Prianka Bhattacharjee}
\email{Prianka.Bhattacharjee@uvm.edu}
\affiliation{\institution{University of Vermont}\city{}\state{}\country{}}

\author{Joseph P. Near}
\email{jnear@uvm.edu}
\affiliation{\institution{University of Vermont}\city{}\state{}\country{}}

\author{Yuanyuan Feng}
\email{Yuanyuan.Feng@uvm.edu}
\affiliation{\institution{University of Vermont}\city{}\state{}\country{}}

\begin{document}

\begin{abstract}
The increasing adoption of differential privacy (DP) leads to public-facing DP deployments by both government agencies and companies.
However, real-world DP deployments often do not fully disclose their privacy guarantees, which vary greatly between deployments. Failure to disclose certain DP parameters can lead to misunderstandings about the strength of the privacy guarantee, undermining the trust in DP.
In this work, we seek to inform future standards for communicating the privacy guarantees of DP deployments. Based on semi-structured interviews with 12 DP experts, we identify important DP parameters to communicate privacy guarantees for transparency and trust. We further elaborate why and how these parameters should be disclosed according to expert consensus.
\end{abstract}
\maketitle

\section{Introduction}

Differential privacy (DP)~\cite{dwork2006calibrating, dwork2014algorithmic}, a mathematical framework for ensuring the privacy of individuals when analyzing sensitive data, has gain traction in real-world adoption. The U.S. Census Bureau~\cite{censusdpuse}, Israel's Ministry of Health~\cite{hod2024differentially}, Google~\cite{googledpuse}, Apple~\cite{appledpuse}, Microsoft~\cite{microsoftdpuse}, LinkedIn~\cite{kenthapadi2018pripearl}, and Facebook~\cite{evans2023statistically} have deployed DP in their datasets or systems to ensure data privacy.

Responsible deployment of DP requires careful disclosure of the specific parameters for the deployment's privacy guarantee. Unfortunately, there is not yet a standard for communicating these parameters, which can vary greatly between deployments. Failure to disclose certain parameters or misleading descriptions of parameter values, can lead to misunderstandings about the strength of the privacy guarantee---even for DP experts. In one example, a confusing statement about the unit of privacy in a DP data release~\cite{bassolas2019hierarchical} led DP researchers to believe the privacy guarantee was weaker than claimed~\cite{houssiau2022difficulty} and required clarification by the original authors~\cite{bassolas2022reply}.

While DP guarantees are mathematically rigorous, their complexity and context-dependence can make them difficult to interpret without sufficient detail. As a result, various stakeholders (e.g., auditors, data analysts, researchers, privacy professionals) need clear documentation of how DP is implemented to evaluate privacy risks and utility trade-offs. For example, analysts must understand privacy parameters to budget queries appropriately, researchers and auditors need to verify whether implementations apply DP meaningfully. Even when simplified, these details can help build public trust by demonstrating the strength and accountability of the system. Our work is motivated by the need to support these audiences with transparent, structured, and accessible DP disclosures.

Prior work on DP communication~\cite{nanayakkara2023chances} has studied \emph{how to explain} one or two important parameters (e.g., the privacy parameter $\epsilon$), without considering \emph{what parameters} should be explained to accurately convey the privacy guarantee rigorously.
Prior work also primarily examined non-expert end-users' understanding, which calls for simplified communication of DP. While simplified formats support accessibility for the general public, they lack precision for technical users or fail to comprehensively convey DP guarantees.
In contrast, our study targets \emph{technical users} and focuses on \emph{what parameters} are important: we seek expert opinions towards a standardized DP label, to accurately and completely communicate privacy guarantees.

Our study aims to answer two key research questions:
\begin{itemize}[leftmargin=12pt,itemsep=1pt,topsep=2pt]
\item \textbf{RQ1:} What parameters should be included in a DP label to communicate privacy guarantees of DP deployments?

\item \textbf{RQ2:} For parameters identified in RQ1, what are the key considerations (e.g. normal ranges, target audiences, methods of presentation) to ensure effective communication?
\end{itemize}
To answer these questions, we conducted a qualitative semi-structured interview study with 12 DP experts to seek expert consensus. We asked experts about the most important parameters (numeric and otherwise) for communicating about DP guarantees; about the normal ranges or values for these parameters; about the suitability of each parameter for communicating with both experts and non-experts; and about how to present parameters to various audiences.

\paragraph{New insights.}
Our results indicate significant consensus about \emph{what} to communicate: experts agreed that important parameters like $\epsilon$, $\delta$, and the unit of privacy are vital for transparency in DP deployments. Compared to the closest related work~\cite{dwork2019differential}, our quantitative analysis enumerates new parameters considered important by experts but not covered in previous work (e.g. deployment model), and additionally describes why each parameter is important, insights about how to set parameter values, challenges associated with communicating about parameters with various audiences, and advice for presenting DP guarantees. Experts in our study raised significant open questions about communicating these parameters to various stakeholders, and there was limited consensus around normal ranges for some parameters. 

\paragraph{Contributions.}
First, we synthesize 12 DP experts' consensus into an essential list of parameters that should be communicated for DP deployments, which lays the foundation for a comprehensive and theoretically sound DP communication standard. Second, we articulate the important properties and considerations for each parameter, which bridges the gap between theoretical guarantees and real-world DP deployments. Third, we discuss key challenges and open questions for communicating privacy guarantees, specifically regarding our proposed DP label format, which contributes to enhancing the transparency and trust in DP deployments.


\section{Related Work}
\subsection{Communicating Differential Privacy}
\label{sec:communicating-dp}
Differential privacy (DP) \cite{ dwork2006calibrating, dwork2014algorithmic} is a formal definition of individual privacy, typically achieved by adding Laplacian or Gaussian noise to computed statistics.
Formally, two datasets \( D \) and \( D' \) are neighbors if they differ by one individual’s data. A mechanism \( M \) satisfies  \((\varepsilon, \delta)\)-DP if, for all neighboring datasets \( D \) and \( D' \) and possible outcomes \( S \), $\Pr[M(D) \in S] \leq e^{\varepsilon} \cdot \Pr[M(D') \in S] + \delta$.
Here, $\varepsilon$ (epsilon) is the privacy parameter (also called the privacy budget). Smaller $\varepsilon$ values yield stronger privacy, and larger values yield weaker privacy. The $\delta$ (delta) parameter relaxes the definition for rare events, and is typically set very small.

Effectively communicating the nuances of DP is challenging. For example, Cummings et al.~\cite{cummings2021need} found textual descriptions ineffective for communicating DP guarantees to end-users.
Many studies with end-users have explored other ways to intuitively convey DP~\cite{bullek2017towards, franzen2022private, wen2023influence, karegar2022exploring, xiong2020towards, xiong2022using, xiong2023exploring, nanayakkara2023chances, ashena2024casual, kuhtreiber2022replication} .
Some studies focused on explaining the privacy risks associated with DP deployments. For example, Franzen et al.~\cite{franzen2022private} found that, although quantitative risk communication formats can improve comprehension, users with low numeracy skills often struggle to grasp the explanations fully.
Similarly, Nanayakkara et al.~\cite{nanayakkara2023chances} tackled the challenge of explaining the epsilon ($\varepsilon$) parameter, which is often abstract and mathematically complex for non-expert users. They proposed using everyday metaphors and simplified risk explanations, to help users grasp how varying epsilon values affect their data privacy.

Other studies examined the effectiveness of visual tools in explaining DP to end-users~\cite{xiong2023exploring, xiong2022using, bullek2017towards, wen2023influence, ashena2024casual, nanayakkara2022visualizing} . For example, Xiong et al.~\cite{xiong2023exploring} found that explanatory illustrations and animations, such as data flow diagrams and noise heat maps, significantly improved user comprehension of privacy-utility trade-offs across central, local, and shuffler DP models.
Karegar et al.~\cite{karegar2022exploring} examined metaphors for explaining DP, such as blurring images. 
They found that metaphors improve DP understanding, but can cause misinterpretation.

In summary, prior work has largely studied \emph{non-experts}, and focused on \emph{how} to communicate DP guarantees.
In contrast, our work studies \emph{DP experts}, and focuses on \emph{what to communicate}.

\subsection{Privacy Labels}
\label{sec:privacy-labels}
Inspired by standardized food nutrition labels, labels have been increasingly adopted to communicate privacy information~\cite{obar2020biggest, kelley2009nutrition}. For instance, Kelley et al.~\cite{kelley2009nutrition, kelley2010standardizing} developed one of the first privacy nutrition labels to visually communicate how organizations collect, use, and share personal data. This simplified complex privacy policies and helped users locate relevant information. 
privacy labels have since been proposed for datasets~\cite{holland2020dataset}, Internet of Things (IoT) devices~\cite{emami2020ask, emami2021informative}, and mobile applications~\cite{cranor2022mobile}. 
However, studies have revealed that Apple's privacy labels for iOS apps can be difficult to locate and sometimes misleading~\cite{zhang2022usable, cranor2022mobile, kollnig2022goodbye}, indicating designing effective privacy labels is challenging and requires rigorous user studies.
Prior work has suggested using labels to explain DP~\cite{smart2024models,  xiong2020effect}, but did not give concrete design proposals.
This work also attempts to draw insights towards designing a comprehensive privacy label for DP, informed by expert opinions.

\subsection{Transparency and Trust in DP}
\label{sec:transparency-dp-deployments} 
Trust in DP deployments depends crucially on transparency.
Closest to our work, Dwork et al.~\cite{dwork2019differential} proposed the \emph{Epsilon Registry}, a communal resource to support documentation of differential privacy deployments by sharing information about $\varepsilon$, design choices, and implementation context. 
They conducted qualitative interviews with 11 individuals from 7 organizations who were involved in DP deployments in 2016.
Different from our study, they focused on specific properties of the participants' deployments, rather than DP transparency in general; they asked participants detailed questions specifically about epsilon (privacy loss) and privacy unit (granularity), but did not ask them to discuss other important parameters.


Our study emphasizes overall transparency in DP deployments and provides both deeper and broader results. First, our interview questions encourage participants to enumerate all important parameters without priming participants to focus on a predetermined set; our results include analysis of roughly twice as many parameters as Dwork et al. Second, our interviews investigated several in-depth aspects not covered by Dwork et al.: (a) \emph{why} each parameter is important; (b) the typical ranges for parameter values; (c) the challenges of communicating each parameter, and the target audience for each; and (d) suggestions for how to communicate each parameter to various audiences. Third, Dwork et al.'s 2016 study drew from several early DP deployments then; participants in our study discuss important issues (e.g.deployment models, utility information, and empirical privacy metrics) that were not as well-understood nearly 10 years ago.

Dwork et al.~\cite{dwork2019differential} inspired several attempts to build actual Epsilon Registries, including the Oblivious Privacy Deployments Registry~\cite{oblivious2024epsilonregistry}, which documents privacy parameters used in real-world DP deployments.
Desfontaines'~\cite{desfontaines2024realworld} informal list of deployments documents additional properties, including the unit of privacy.
The Wikimedia Foundation~\cite{wikimedia2024pageviews} publishes DP page-view datasets, with documentation of the mechanisms used and pre- and post-processing procedures.
Despite these efforts, gaps remain in how transparency is achieved for DP deployments, particularly in communicating critical parameters such as the unit of privacy, which plays a key role in defining privacy guarantees. 
Although some registries, such as the Oblivious/OpenDP registry~\cite{oblivious2024epsilonregistry}, include relevant information under fields such as "scope", they do not always explicitly label or emphasize the unit of privacy. 

Moreover, transparency can be counterproductive without clear communication. Cummings et al.~\cite{cummings2023centering} emphasized that incomplete or unclear descriptions of DP can erode user trust.
Building on these efforts, our work aims to identify \emph{all} key parameters for responsible transparency in DP deployments. Our DP label informs future epsilon registries by avoiding transparency traps~\cite{cummings2023centering}, such as focusing too heavily on certain parameters while overlooking broader privacy considerations.

\subsection{Transparency with DP Stakeholders}

Our study aims to elicit how to ensure transparency from expert opinions; in practice, DP deployments impact many different stakeholders with differing levels of expertise and different requirements for transparency. Here, we provide a high-level overview of prior work associated with different groups of stakeholders, how our results might apply to those groups, and the future research necessary to ensure transparency for each.

\paragraph{End-users.}
As described earlier, most studies investigating \emph{how} to communicate about DP have focused on end-users---members of the general public with no expertise in DP---and the results suggest that communicating DP guarantees with this audience remains challenging. Our results on transparency are likely applicable to this group, but significant additional research is required to determine \emph{how} to communicate with end-users.

\paragraph{Data users.}
Users of DP data are primarily concerned with the utility of DP data, rather than its privacy properties, influencing their expectations for transparency. Moreover, data users are typically not DP experts, which complicates DP communication efforts. To date, only a few studies have investigated communication of DP properties with data users.
Williams et al.~\cite{williams2024disclosing} investigated the utility requirements of economists via a survey; they found that economists have low awareness of DP, are not convinced it is needed, and are not sure how to decide whether a data release provides sufficient utility for downstream tasks.
These results align with the response to the US Census Bureau's use of DP in the 2020 Decennial Census~\cite{abowd2018us}. Users of Census data expressed concern over the necessity for DP and its impact on data quality~\cite{mueller20222020, ruggles2019differential, kenny2021use}.
Other work has focused on measuring utility for specific tasks, rather than communicating with data users about utility. For example, Rosenblatt et al.~\cite{rosenblatt2023epistemic} study whether results from population-level studies would be reproducible if DP synthetic data were used; Hay et al.~\cite{hay2016principled} compare the accuracy of DP algorithms on a variety of datasets for specific histogram tasks.
Experts in our study included elements of utility in their recommendations for transparency, but were not primarily focused on the needs of data users. The limited existing work in this area suggests that additional research on how to communicate with DP data users to achieve transparency is needed.

\paragraph{Legal and regulatory stakeholders.}
As described by Garrido et al.~\cite{garrido2022lessons}, legal and regulatory compliance officers represent a key group of stakeholders in transparency for DP guarantees. A significant amount of prior work has examined the relationships between DP and legal requirements for privacy
Nissim and Wood~\cite{nissim2018privacy} and Sokolovska and Kocarev~\cite{sokolovska2018integrating}
developed frameworks relating legal privacy requirements with technical requirements, and analyze the extent to which DP's technical properties satisfy these requirements. 
Holzel~\cite{holzel2019differential} and Cohen and Nissim~\cite{cohen2020towards} perform similar analyses, but specifically for GDPR. 
Chin and Klienfelter~\cite{chin2011differential} and Cohen et al.~\cite{cohen2022private} analyze specific use cases (Facebook ads and the US Decennial Census, respectively).
The body of prior work suggests that DP can play an important role in fulfilling legal and regulatory privacy requirements, but that (1) the specifics details of the DP guarantee is important for fulfilling legal requirements, and (2) additional context beyond the DP guarantee itself is also important to ensure compliance.
Our transparency results align with the technical features of DP considered important for legal compliance in prior work, suggesting that transparency for DP experts may overlap considerably with transparency for legal experts. Additional research with legal and regulatory experts is needed to identify gaps between these requirements and to investigate how DP guarantees should be communicated with this group.


\section{Method}
We conducted a qualitative interview study with 12 DP experts.
Our semi-structured interview design was motivated by the open-ended nature of our research questions. One of our goals was to identify the important parameters for transparency in DP deployments, and key considerations associated with them; an open-ended, qualitative study is best suited for answering these questions~\cite{oates2022researching}.
This section details the participant recruitment, the interview procedures, the data analysis process, and ethical considerations of our study.

\subsection{Sample \& Recruitment}
\label{sec:sample-recruitment}

\paragraph{Recruitment:}
We used the purposive sampling method to recruit DP experts who are at least 18 years old and have a minimum of five years of research experience or practical involvement with DP. 
 Given the niche expertise required for this study, we first approached potential qualified participants via targeted recruitment emails sent to our professional network and the OpenDP mailing list (where DP experts gather).  We also used snowball sampling to seek peer recommendations of additional qualified candidates from enrolled participants.
 We successfully recruited ten experts via targeted email and two additional experts through peer recommendation.
Our study invitation email includes an eligibility survey (Appendix \hyperref[app:appendixA]{A}) focused on their professional background in DP.
Our recruitment method and sample size align with prior qualitative studies in DP area~\cite{dwork2019differential, sarathy2023don, garrido2022lessons}. Similarly, we were limited in sample size by the challenge of recruiting experts in DP. See Section~\ref{sec:limit-future-work} for additional discussion of this limitation.

\paragraph{Our sample:}
12 DP experts from 10 different organizations participated in our study between May and August 2024. 
On average, these experts had 7.5 years of experience in DP. We assigned participant numbers P01--P12 to these experts. 
Table~\ref{tab:expert_demographics_n=12} details the expert demographics in terms of work experience, job roles, and gender, as well as the industries they are engaged with.
See Table {\ref{tab:expert_demographics} in Appendix \hyperref[app:appendixC]{C} for additional expert demographics.

Our participants include two professors and one postdoc at research-focused universities whose primary publication area is DP; five individuals who have played key roles in the design or development of US government DP deployments; and six individuals who have played key roles in the design or development of industry DP deployments. All are currently recognized experts in the area of DP. Where available, we list the H-index of each participant in Table~\ref{tab:expert_demographics_n=12} as an (imperfect) proxy for expertise.

\begin{table}
\centering
\begin{tabular}{p{.4cm}|p{1.0cm}|p{2.85cm}|p{.7cm}|p{.8cm}|p{.9cm}}

\hline
\rowcolor[HTML]{D3D3D3} 
\textbf{ID} & \makecell{\textbf{Exper-} \\ \textbf{ience}\\ \textbf{(Years)}} & \textbf{Role} & \makecell{\textbf{Gen-} \\ \textbf{der}} & \makecell{\textbf{PhD} \\ \textbf{(Y/N)}} & \makecell{\textbf{H-} \\ \textbf{index}} \\
\hline
P01 & 10+ & Academic Researcher  & F & Yes & 25-30 \\
P02 & 5 - 10 & Academic Researcher & M & No & 30+ \\
P03 & 10+ & Data Scientist & F & Yes & 5-10 \\
P04 & 5 & Data Privacy Engineer & M & No & N/A \\
P05 & 5 & Academic Researcher & F & Yes & 5-10 \\
P06 & 5 & Academic Researcher & F & Yes & 25-30 \\
P07 & 5 - 10 & Data Privacy Engineer & M & Yes & 15-20 \\
P08 & 5 - 10 & Data Scientist & F & Yes & 10-15 \\
P09 & 5 & Data Privacy Engineer & M & No & 5-10 \\
P10 & 5 - 10 & Software Engineer Building DP Tools & M & No & N/A \\
P11 & 5 - 10 & Data Scientist  & M & No & 5-10 \\
P12 & 5 - 10 & Data Privacy Engineer & M & No & 5-10 \\
\hline
\end{tabular}
\caption{Expert Demographics, $n = 12$}
\label{tab:expert_demographics_n=12}
\end{table}

\subsection{Interview Procedure}
\label{sec:interview-procedure}

The research team informed participants about the study's data collection, processing, and storage procedures, obtained their consent to participate, and assessed their eligibility for the study via an eligibility survey (details in Appendix \hyperref[app:appendixA]{A} ). The research team conducted the interviews via Microsoft Teams and made audio recordings after obtaining verbal consent from the experts. 
The semi-structured interviews followed a three-part format:
\begin{enumerate}[leftmargin=14pt, itemsep=0pt,topsep=3pt] 
    \item The first part of the interview focused on identifying key differential privacy (DP) parameters that experts considered essential for inclusion in DP systems and deployments. Experts were asked to elaborate on the significance of each parameter and whether specific parameters are often overlooked or underestimated in DP discussions.
    \item The second part of the interview explored why each parameter should be included in our proposed DP label, examining its typical or "normal" range within the context of DP, its relevance to the general and technical audiences. Experts were also asked to assess the applicability and effectiveness of these ranges, providing examples where necessary.
    \item The third part of the interview explored the design, structure, and presentation of the DP label. We seek experts' feedback on our proposal for a layered design, as well as their insights on how each parameter should be displayed, where the labels should be placed for maximum visibility, and other recommendations for improving the label's layout and accessibility.
\end{enumerate}
\noindent To support a shared understanding of the study’s framing, we began each interview by presenting visual examples of nutrition-style labels (e.g., food and iOS privacy labels) and explained our goal of developing a similar label for DP deployments. We clarified that the term “parameter” was defined broadly as: any feature, metric, or concept that experts believed was important for understanding or communicating a DP guarantee. This included, but was not limited to, numeric values like $\varepsilon$ or $\delta$. Participants were encouraged to consider contextual and structural elements (e.g., the unit of privacy, privacy loss budgeting, or utility constraints), as well as any other information they deemed critical for interpreting a DP deployment. 

To ensure clarity of the questions and estimate the duration of the interview, we piloted the interview questions piloted with two graduate students in our research group who were familiar with DP not involved in the study design. After piloting, we reduced the number of questions and adjusted the order of some questions. The finalized interview guide appears in
(Appendix \hyperref[app:appendixB]{B}).

\subsection{Data Analysis}
\label{sec:data-analysis}
To analyze the interview transcripts, we used a hybrid thematic analysis process~\cite{proudfoot2023inductive, fereday2006demonstrating, braun2006using}
 to systematically identify recurring themes related to our research questions. This approach combined deductive coding, guided by predefined DP concepts, with inductive coding to allow new themes to emerge from the data, ensuring that both established DP principles and novel insights were captured.

The interviews were transcribed using Microsoft Teams' automated transcription feature and carefully reviewed and corrected to ensure accuracy and completeness.
Once familiarized with the data through repeated readings, two researchers independently open coded the 12 transcripts.  Although the predefined DP concepts informed the researcher's understanding, the coding process prioritized emergent themes from the transcript contents.

Following the independent coding phase, the two researchers collaborated to merge their individual codebooks into a combined version. Where disagreements arose, they discussed the differences and, if necessary, sought input from the broader research team to reach a consensus. Once a unified codebook was finalized for each transcript, the researchers grouped the codes by parameter and synthesized them into broader themes. This process allowed for systematic organization of the data while preserving its complexity and nuance. Finally, the researchers revisited the data to validate the themes, ensuring that they accurately reflected the content of the interview. This collaborative and iterative approach ensured the analysis was rigorous, coherent, and faithfully represented the insights provided by the experts.

\subsection{Ethical Considerations}
\label{sec:ethical-considerations}
This study received approval from our university’s Institutional Review Board (IRB).
All participants received information about the study’s objectives, data collection, processing, and storage procedures and provided their consent through the eligibility survey (Appendix \hyperref[app:appendixA]{A}). They also verbally reaffirmed their consent at the start of each interview.
To protect participants’ privacy, only general demographic information and professional background were collected, and identifiable data were anonymized during transcription.
Audio recordings from the interviews were transcribed, coded with unique identifiers, and securely stored.
All participants received equal compensation of \$40 in the form of an electronic gift card upon completing the interview. Four experts chose to donate their compensation to charity.

\section{Results} \label{sec:results}
To answer RQ1, we present the key DP parameters that experts deemed essential for communicating privacy guarantees, and synthesize why they are important.(Section~\ref{sec:results:parameters}). Then, we answer RQ2 by presenting our qualitative analysis results of the communication challenges (Section~\ref{sec:communication_challenges}), typical ranges for parameter values (Section~\ref{sec:parameter_ranges}), audience relevance (Section~\ref{sec:Target audience}), and recommended presentation formats for these parameters (Section~\ref{sec:parameter_presentation}). 

\begin{table}
\vspace*{-10pt}
    \centering %
    \begin{tabular}{|p{6.5cm}|c|}
    \hline
    \textbf{Parameter} & \textbf{\S}\\ \hline
    Privacy parameters & \ref{sec:parameters}\\
    Unit of privacy & \ref{sec:unit of privacy}\\
    Utility information & \ref{sec:utility information}\\
    Mechanism used & \ref{sec:mechanism used}\\
    Algorithm hyperparameters & \ref{sec:algorithm hyperparameters}\\
    Deployment model & \ref{sec:deployment model}\\
    Empirical privacy metrics & \ref{sec:empirical privacy metrics}\\
    Privacy interpretation-semantic & \ref{sec:privacy interpretation or semantics}\\
    Other parameters & \ref{sec:other parameters}\\
       
    \hline
    \end{tabular}
    \caption{Parameters Mentioned by Experts}%
    \label{tab:1} %
    \vspace*{-20pt}
\end{table}

\subsection{Important DP Parameters} \label{sec:results:parameters}

We identified nine key categories of parameters that experts considered essential for defining and communicating DP guarantees: privacy parameters (e.g., epsilon, delta), unit of privacy, utility information, mechanism used, algorithm hyperparameters, deployment model, empirical privacy metrics, privacy interpretations or semantics, and other parameters.

Table~\ref{tab:1} presents these categories, which serve as an organizing structure for the wide range of parameters discussed during the interviews. A comprehensive list of individual parameters and how they map to these categories is included in Appendix~\hyperref[app:appendixD]{D}.
In the following subsections, we elaborate on each parameter category, focusing on its role in enhancing transparency in DP deployments.

\subsubsection{Privacy Parameters}
\label{sec:parameters}
Privacy parameters are core metrics in the DP literature for defining the strength of privacy guarantees, balancing privacy protection and data utility \cite{dwork2006calibrating, dwork2014algorithmic}. DP experts in our study agreed on three key privacy parameters: Epsilon ($\epsilon$), Delta ($\delta$), and other privacy loss measures (such as rho ($\rho$) for zCDP~\cite{bun2016concentrated}).

The privacy parameter assigned to a data release is often referred to as the privacy budget \cite{wood2018differential}, which quantifies the allowable privacy loss in a system. In practice, parameters like epsilon and delta are central to real-world DP deployments and disclosed in the US Census Bureau’s 2020 data release~\cite{abowd20222020} and other industry deployments~\cite{apple2017learning, fanti2015building}. Experts in our study unanimously emphasized the need to include privacy parameters in a DP label for transparency.
As one expert remarked, \textit{“These are the things that's gonna give experts a good understanding of the'real' privacy guarantees of the system.”}

\paragraph{Epsilon ($\epsilon$)}
\label{sec:epsilon}

\noindent Epsilon ($\epsilon$) is the cornerstone parameter in DP \cite{dwork2006calibrating}, quantifying the indistinguishability between datasets and measuring privacy strength. It is the tuning knob for balancing privacy and accuracy as it determines how much noise to be introduced \cite{wood2018differential}. Smaller values provide stronger privacy but reduced utility \cite{near2023guidelines} making epsilon essential for evaluating DP systems, and defining privacy-utility tradeoffs.

\paragraph{Direct Impact on Privacy Guarantees:} Eight experts highlighted epsilon as a key parameter directly impacting privacy guarantees. P01 stated, \textit{“Epsilon is the main parameter that controls the privacy guarantee. So I think it's like really important, like at a high-level value”}. Another described espsilon as a \textit{“knob that you can tune, and it's sort of the center of a lot of decision-making that happens”} (P05), highlighting its central role in DP systems.

\paragraph{Enabling Interpretation and Comparability:}
Five experts highlighted epsilon’s role in interpreting and comparing DP systems.
One expert remarked, \textit{"...it gives you some way of comparing things"} (P07). Another expert described epsilon as \textit{"a Rosetta Stone across a whole bunch of different kinds of differential privacy"} (P09), underscoring its importance in bridging various DP frameworks.

\paragraph{Delta ($\delta$)}
\label{sec:delta}

\noindent Delta (\( \delta \)) complements epsilon (\( \epsilon \)) by bounding the probability of privacy failure in worst-case scenarios. While $\epsilon$ quantifies indistinguishability, $\delta$ accounts for residual risks, offering a fuller picture of a DP mechanism’s privacy guarantees \cite{near2021programming}.

\paragraph{Ensuring Robust Privacy Guarantees:}  
Three experts emphasized delta’s role in capturing worst-case privacy risks. P01 explained, \textit{"...in the worst case, delta captures how much your data could be exposed in the clear."}  Another noted, \textit{“Epsilon and delta together represent the degree of privacy being offered”} (P03).
 
\paragraph{Prevents Privacy Theatre:}  
Two experts stated that delta safeguards against weak or misleading DP deployments. P01 explained, \textit{“you should include delta to prevent against malicious uses, like people hiding things they are doing”}. P05 warned, \textit{“if you have a delta that’s super big, then it’s just a poor deployment...not offering the protections that people think they’re getting.”}   
Setting an explicit bound on privacy failure ensures transparency and upholds DP's credibility.

\paragraph{Other Privacy Loss Measures}
\label{sec:other_privacy_loss_measures}

\noindent Beyond epsilon ($\epsilon$) and delta ($\delta$), alternative frameworks like $\rho$-zero concentrated DP~\cite{bun2016concentrated} and Rényi DP \cite{mironov2017renyi} can offer improved privacy loss quantification.

\paragraph{Contextualizing Privacy Guarantees:}  
P01 and P07 emphasized that incorporating alternative privacy loss measures improves the accuracy and robustness of privacy guarantees. P01 explained, \textit{"If you stick to epsilon as the primary parameter, composition under frameworks like rho or Rényi DP often results in a looser bound... Including these parameters ensures the guarantees are fully contextualized."}

\paragraph{Improving Privacy Loss Accounting:}
Several experts noted that alternative privacy loss measures help in quantifying privacy loss across multiple data releases. P08 stated, \textit{"You can quantify the privacy loss... and compose everything to still understand the combined privacy loss."} By providing additional granularity, these parameters enhance the interpretability of privacy guarantees for technical stakeholders.

\paragraph{Enhancing Transparency and Trust:} 
P04 and P11 emphasized that including these parameters enhances transparency and strengthens trust in DP systems. P04 remarked, \textit{"Estimates of privacy loss parameters should be included... If you don’t include those, it’s either defeating the purpose or being dishonest."} This underscores their role in ensuring the integrity of privacy guarantees and fostering confidence in DP deployments.

\subsubsection{Unit of Privacy}
\label{sec:unit of privacy}
The unit of privacy defines what and who is being protected in a DP system, shaping the scope and granularity of privacy guarantees \cite{near2023guidelines}. It varies across applications—ranging from record-level (or event level) to user-level privacy—and influences both protection levels and utility trade-offs. Experts discussed its role in ensuring transparency, optimizing utility, and the challenges of interpretation.

\paragraph{Defines Scope of Protection and Ensures Transparency:} Eight experts emphasized that specifying the unit of privacy clarifies the scope of protection. P03 noted, \textit{"Unit of privacy ... tells you what the guarantee applies to"}.
P04 reinforced this, stating, \textit{" DP is giving a privacy guarantee. You should know who the guarantee applies to. The privacy guarantee is meaningless if you don't have that."} 
Explicitly defining it also pushes organizations to clarify their privacy commitments, as P07 pointed out: \textit{"It forces companies to actually think about it... It would push companies to say things like, 'We do user-level DP,' but what we actually mean is device-level DP."} This transparency helps prevent "privacy theater" and ensures that the stated guarantees align with actual protections.

\paragraph{Utility Implications:}
The unit of privacy helps optimize DP mechanisms for practical use. P10 described it as \textit{"a very useful tool as an algorithm designer to tune utility."} 
By specifying the granularity of privacy, that is, privacy at different levels (e.g., user, session), organizations can tailor mechanisms to provide robust protections while maintaining usability and utility. This granularity ensures that privacy guarantees are not only meaningful but also adaptable to the specific needs of different applications.

\subsubsection{Utility Information}
\label{sec:utility information}
Utility information emerged as essential for DP deployments, encompassing metrics related to data accuracy after privacy mechanisms are applied. Experts highlighted its role in balancing privacy guarantees with meaningful data use.

\paragraph{Ensures Transparency:} Utility metrics help prevent overly noisy or misleading outputs. P03 noted, \textit{"It keeps algorithm developers honest... forcing them to analyze their results so they're not producing wildly noisy, problematic data."} 

\paragraph{Influences Incentive to Share:} Some experts emphasized utility’s role in shaping individuals' willingness to share data. P05 stated, \textit{"there's actually a pretty strong case for including it even to data subjects... it influences how they feel about the decision to share their data or how they feel about the privacy that they're being offered"}

\paragraph{Facilitates Evaluation of Data Quality:} Utility metrics, such as error rates, are crucial for evaluating data quality. P09 explained, \textit{"These things are key for an analyst who wants to understand the actual data quality and how close they can assume it will be to the underlying dataset"}. P08 emphasized that such metrics are \textit{"super nice for the data user"}.

\subsubsection{Mechanism Used} 
\label{sec:mechanism used}
The mechanism in DP dictates how privacy guarantees are implemented \cite{near2023guidelines}, shaping both precision and interpretability. It determines the way in which noise is applied, data is synthesized, or queries are answered, affecting the reliability of privacy assurances.

\paragraph{Precision of Guarantee Definition:}
The choice of mechanism defines privacy guarantees based on statistical properties and use cases. P01 noted, \textit{"Epsilon, delta may not bind as tightly for every mechanism. The nature of the guarantees in a more nuanced way depends on the mechanism that is used."} 
P12 connected mechanisms' role to their ability to provide statistically rigorous definitions of privacy stating, \textit{"This is directly required to actually come up with a bound on the statistical power of a test."}

\paragraph{Enables Interpretation and/or Comparison:} Mechanisms facilitate understanding and comparison of privacy guarantees across different deployments. P10 explained, \textit{“You sort of need to know what the privacy definition is to interpret the parameters. It’s helpful to characterize mechanisms in ways that don’t require providing the full mechanism details”}. This supports technical users in evaluating trade-offs.

\subsubsection{Algorithm Hyperparameters} 
\label{sec:algorithm hyperparameters} 
Algorithm hyperparameters are crucial in differential privacy (DP) implementations because these parameters, such as clipping bounds, noise multipliers, and batch sizes, should be included in transparency efforts like a DP label to inform stakeholders about privacy and data quality.

\paragraph{Transparency:} Hyperparameters impact data usefulness and assess result quality. As P04 noted, \textit{“It tells you how confident you should be in the results... Putting those in context of how good the release is, is also important”}. P08 added, \textit{“Including clipping or clamping parameters ensures that the data was appropriately preprocessed, reducing bias and enhancing trust”}.

\paragraph{Understanding Utility:} Hyperparameters clarify the privacy-utility trade-off. P06 explained, \textit{“Including these details would allow a better understanding of how much signal you are getting, and how much of it is clipped, which in turn impacts the utility”}.

\subsubsection{Deployment Model}
\label{sec:deployment model}
The deployment model determines how data is processed, stored, and accessed, influencing privacy and security. It can involve local processing on a user's device (as in local DP), centralized processing by a data collector (as in central DP), or decentralized models like federated systems. Each model presents different privacy and security trade-offs, making it crucial to assess the privacy protections at various stages.

\paragraph{Significant Impact on Threat Models:}
The deployment model directly affects the threat model and associated security and privacy guarantees in a deployment. As P05 noted, \textit{"It implies significantly different threat models, which impact data security concerns"}. Local DP avoids the need for a trusted data collector and reduces the risk of data breaches, while central DP increases the risk of a data breach but offers better utility.

\subsubsection{Empirical Privacy Metrics}
\label{sec:empirical privacy metrics}
Empirical privacy metrics measure privacy experimentally, via reconstruction, re-identification, membership inference, or other attacks, or via experimental auditing procedures~\cite{jagielski2020auditing,carlini2022membership,nasr2018machine}. The results can provide a lower bound or average-case approximation of the privacy protection provided by a data release. One or many empirical metrics can be provided based on the concerns for a specific dataset. 

\paragraph{Practical Risk Assessment:}
Empirical privacy metrics help assess actual attack risks. As P06 stated, \textit{"I think empirical results are just always good to have for anyone because that's the thing that is tangible for people and people can assess their risks."}  Empirical metrics are broadly applicable to evaluate datasets with or without DP, and may be easier for non-experts to understand, so they may be particularly useful for end-users, data users, and policymakers.

\subsubsection{Privacy Interpretation or Semantics}
\label{sec:privacy interpretation or semantics}
Privacy interpretation or semantics communicate privacy by directly stating the semantics of the guarantee, often as a bound on the trade-off between Type 1 and Type 2 error in a hypothesis test. 

\paragraph{Practical Risk Assessment:} 
The interpretations are important for conveying privacy risks clearly, particularly for decision-makers both experts and non-experts who are familiar with concepts like type 1 \& 2 errors better grasp privacy risks in practice. As P11 noted, \textit{"I think it helps ... to have something like a type 1, type 2 error interpretation that is closer to a plain language statement..."}

\subsubsection{Other Aspects or Parameters}
\label{sec:other parameters}
Experts highlighted additional aspects or parameters that, while mentioned less frequently, could reflect nuanced dimensions of DP to improve transparency, facilitate interpretation, and support informed decision-making.

These include the \textbf{data creation pipeline/contextual integrity}, \textbf{data release metadata}, \textbf{goals of the release}, \textbf{DP software tool used} (e.g., OpenDP) \textbf{implementation and vetting details,} (e.g., compliance with DP principles and verifying correctness), \textbf{average-case guarantees}, \textbf{guidance for normal ranges of parameters}, \textbf{group privacy information} and \textbf{public information used}. Notably, P09 emphasized the importance of documenting the data creation pipeline and release goals stating, \textit{"Error in DP often comes from data governance rather than aggregation... Ensuring clear documentation and attribution of all steps is critical."} This perspective is derived from this expert's prior experience with DP deployments. 

Although these aspects were mentioned less frequently, they may become increasingly important as DP deployments grow in scale and complexity. In particular, many of these suggestions reflect the principle of \textit{contextual integrity}~\cite{nissenbaum2009privacy, nissenbaum2004privacy}—the idea that privacy depends not only on data protection mechanisms but also on preserving appropriate information flows relative to the context in which data is collected, processed, and shared. Capturing metadata about the release context, tool choices, and intended use aligns with this framework and supports transparency grounded in the expectations and norms of the data environment.

\subsubsection{Overlooked Parameters} \label{sec:results:overlooked}
Many experts noted that discussions on DP tend to overemphasize epsilon and neglect other critical parameters. As P01 remarked, \textit{"Everything besides epsilon is overlooked... but there’s so much more that is important."}

Seven experts highlighted the  \textbf{unit of privacy} as another often-overlooked parameter in DP. P08 observed, \textit{"In discussions, I feel like the unit of privacy is not so much discussed, at least not as much as privacy loss guarantees."} Similarly, P05 remarked, \textit{"I don’t think it’s overlooked in the DP community, but I think that it gets overlooked in documentation around DP deployments."}

Four experts identified \textbf{algorithm hyperparameters} and \textbf{utility metrics} as underestimated parameters that significantly affect DP effectiveness and usability. P06 explained, \textit{"When training with DP, hyperparameters like learning rates and batch sizes become nonlinear and more complex, which is often overlooked."} P03 emphasized, \textit{"All the little algorithm quirks and data properties that impact utility are overlooked... Things that affect any machine learning algorithm probably get overlooked here too."}

One expert noted \textbf{sub-budgets} as often overlooked. P11 pointed out, \textit{"The difference between per-attribute guarantees and the global budget is significant, yet these sub-budgets are rarely discussed."} Similarly, one expert identified \textbf{data universe assumptions} which define the set of possible records in a DP framework, as critical yet often overlooked. P12 remarked \textit{"The definition of universes is probably one of the most overlooked but critical aspects."}

Experts emphasized the need for greater attention to these overlooked parameters to adopt a holistic approach to DP and improve transparency. This would enable stakeholders to fully understand the scope of privacy guarantees and associated trade-offs.

\subsection{Challenges of Communicating Parameters}
\label{sec:communication_challenges}

While many DP parameters are essential for defining privacy guarantees, experts in our study identified significant challenges in making these parameters understandable and useful---particularly for non-experts. We synthesized these overarching challenges in communicating DP parameters below.

\paragraph{Too Technical and Difficult to Understand:}
Experts emphasized that many DP parameters—particularly ($\epsilon$), ($\delta$) and advanced constructs like Rényi DP—are too technical for non-technical audiences. P09 described this as DP’s \textit{“main Achilles heel... you have to have some mathy understanding of it to begin with.”} 
P01  stated that for rho, \textit{"It’s overly technical, and you wouldn’t want to explain this to a general audience."} 
This theme extended to the unit of privacy, deployment models, mechanisms, privacy semantic interpretations, and algorithm hyperparameters, which several experts considered too intricate to be broadly interpretable.

Five experts noted the difficulty in interpreting and comparing privacy units, particularly in complex settings like DPML. P09 explained, \textit{"... If you're doing DPSGD, the privacy unit is harder to describe in human understandable terms."} 

Similarly, experts noted that the privacy semantic interpretations are too complex to understand, making them not useful in practice, as many users are unlikely to invest the time to fully grasp them.
Furthermore, experts also cautioned that mechanism details are too complex for a non-technical audience. P08 added, \textit{"It makes sense for engineers but not if you're sharing it with everybody’s grandma."} This complexity risks alienating general audiences.

For algorithm hyperparameters, experts warned that too many technical details can overwhelm non-experts. As P06 said, \textit{“If they are ML like LLM practitioners but are not laypersons... they would care because just the size of the gradient tells you how much signal you’re getting.”}
In addition, explaining deployment models to non-technical audiences can be challenging, as the nuances between models may be hard to convey clearly. Experts pointed out concerns about information overload when presenting multiple models without prioritization, making the disclosure technical and difficult to understand.

\paragraph{Risk of Misinterpretation and Overemphasis on Numbers:} Several experts raised concerns about overly focusing on privacy parameters and numeric values, which can lead to misleading comparisons. P07 explained, \textit{"It’s much easier to compare numbers... and people would do that and get maybe wrong conclusions.”} P05 added that \textit{"Discussions around DP just became about epsilon... foreclosing all those other issues.”} Without clear interpretive context, these values may obscure rather than clarify privacy risks.

\paragraph{Information Overload and Redundancy:}
Experts warned that including too many technical details could overwhelm users. P07 suggested that advanced parameters be included only in supplementary materials: \textit{"That information benefits only experts, who can check the white paper for details."} P05 observed that parameters like delta are often already set by consensus and may not require user-facing disclosure. \textit{"There’s more consensus around delta, so maybe including it is just giving someone more information than they need.} Other experts highlighted the redundancy of presenting algorithm hyperparameters when the overall privacy guarantee is already disclosed. As P08 put it, \textit{“There is already enough information through the privacy guarantee itself, so disclosing clipping bounds again might be redundant.”}

\paragraph{Limited Relevance and Potential for Harm:}
Utility information presented its own challenges. Some experts argued it has limited value for privacy-conscious users or policymakers. P09 remarked, \textit{"I don't think it's that important for someone who cares about their privacy."} P03 warned that revealing data thresholds could discourage participation: stating, \textit{"If I'm rare enough, I might not fill this out at all because it already told me it's pointless."} Other experts cautioned that organizations could misuse utility metrics to justify weaker privacy protections. P05 noted, \textit{"I wonder if organizations would use it to manipulate people... saying, 'Well, yeah, you have to give up your privacy for better utility."}

\paragraph{Lack of Standards and Complexity of Implementation:}
Experts considered utility information and empirical privacy metrics difficult to standardize or measure. P10 stated that utility information \textit{"is difficult to characterize and will vary greatly from dataset to dataset."}. Experts noted that although it is nice to have empirical privacy metrics to provide a sense of practicality, there is no standard measure available to do so that addresses all user concerns from every perspective. P11 expressed concern about 'privacy theater', since empirical metrics often focus on average-case performance, potentially downplaying worst-case attack scenarios. \textit{"Even if all of your per-attribute privacy budgets are big, every theoretical guarantee you can provide is almost trivial. Then you need empirical attacks to show that you did something useful."} This highlights the need for careful interpretation of results. Experts also noted that it is infeasible to implement and describe all possible attacks in a way that caters to concerns of all the stakeholders.  Metrics that are not on the list can still show up in a way that becomes a source of concern for privacy.

\paragraph{Uncertain Practical Utility:}
Some experts questioned the real-world utility of privacy parameters. P11 remarked, \textit{"In practice, it’s close to useless... we end up with very large values ...for which the guarantees are also not comforting."} This highlights the gap between theoretical guarantees and practical privacy risks.

\subsection{Typical Ranges for Parameters}
\label{sec:parameter_ranges}
This subsection summarizes expert perspectives on typical or context-appropriate ranges for DP parameters, when applicable. Experts emphasized that parameter values vary widely depending on factors like data sensitivity, dataset size, and use case, and that many parameters do not have a standard range. For a subset of parameters, they offered ranges supported by community norms. Because each parameter involves different trade-offs, we present ranges on a parameter-by-parameter basis, focusing only on those for which experts provided concrete numerical guidance. Parameters such as the deployment model or mechanism are not included here, as they do not have ranges. Table~\ref{tab:dp_ranges} summarizes expert-suggested value ranges for key DP parameters.

\paragraph{Epsilon:}
Experts agreed that $\epsilon$ should be context-dependent, with common values ranging from \textbf{0.001 to 20}. Many (P01, P04, P05, P09, P11) recommended $\epsilon \leq 4$ as ideal, though higher values are common in industry. P04 explained, \textit{"Guidance suggests 1 or 0.5, but we've seen higher in industry."} P03 and P06 viewed values up to 10 as reasonable; P06 noted, \textit{"Below 1 means high privacy... above 10 is basically no privacy."} For high-dimensional data, P08 found values up to 20 acceptable.

\paragraph{Delta:}
Experts agreed $\delta$ should be \textbf{very small}, often $10^{-5}$ to $10^{-6}$ (P01). P03 suggested $1/\sqrt{n}$; P08 preferred $1/(\log n \cdot n)$. P05 favored extremely small values to ensure strong guarantees. P01 also noted that larger datasets may demand even smaller deltas for meaningful privacy protection.

\paragraph{Unit of Privacy:}
Most experts supported user-level or individual-level units (P04, P08, P09, P10). P08 noted, \textit{"A normal range is anything that is individual level or a proxy for an individual."} Others (P03, P05) emphasized the need for stakeholder-specific definitions. P09 proposed time-based units like \textit{"one day to one month or a quarter."}

\paragraph{Utility Information:}
Utility needs varied by use case. P03 linked it to group sizes: \textit{"100 is reasonable, 50 is good, 10 is problematic."} P08 stressed that utility depends on context. P09 recommended relative error $\leq$ 5--6\% to avoid misleading data.

\paragraph{Algorithm Hyperparameters:}
Experts had diverse views. P04 emphasized balancing bias and variance; P06 cited clipping norms from 0.1 to 1. P09 highlighted output thresholds for large datasets.

\paragraph{Empirical Privacy Metrics:}
P06 recommended percentiles and averages for risk assessment: \textit{"Gives a clearer picture."} But cautioned that averages may miss individual risks.

\paragraph{Privacy Interpretations or Semantics:}
For privacy interpretations, experts referenced Type 1 error rates (0.01, 0.05, 0.1) and Type 2 power ranges (0.1 to 0.7). These offer intuitive, bounded guarantees, but may not apply when data is already public or low-risk.

\begin{table}
\centering
\small
\begin{tabular}{p{3.5cm}p{4.2cm}}
\toprule
\textbf{Parameter} & \textbf{Typical Range / Guidance} \\
\midrule
Epsilon & $0.001 - 20$ (ideal $\leq 4$) \\
\hline
Delta & \(10^{-5}\) to \(10^{-6}\); or $ \frac{1}{\sqrt{n}} $ \\
\hline
Unit of Privacy & User-level; time-based (day–quarter) \\
\hline
Utility Info & Group size $\geq 50-100$; error $\leq 5-6\%$ \\
\hline
Algorithm Hyperparameters & Clip norm: 0.1–1; tuned noise scale \\
\hline
Empirical Metrics & Percentiles, std, min/max risk \\
\hline
Privacy Semantics & Type I error: 0.01–0.1; Power: 0.1–0.7 \\
\bottomrule
\end{tabular}
\caption{Examples of expert-recommended typical ranges for DP parameters.}
\label{tab:dp_ranges}
\end{table}

\paragraph{Epsilon:}
Experts agreed that epsilon values {vary} by application, data sensitivity, and risk tolerance, with suggested thresholds from {0.001 to 20.} Five experts (P01, P04, P05, P09, P11) considered $\epsilon \leq 4$ common or ideal, though industry settings often use larger values. P04 explained, \textit{"...general guidance suggests keeping epsilon at most 1 or 0.5, but it varies with data sensitivity...we've seen higher values in industry."}
For flexible applications, two experts (P03, P06) found values up to 10 reasonable. P06 noted, \textit{"Below 1 means high privacy, 1 to 10 is medium, but above 10 is basically no privacy."} 
For high-dimensional or synthetic data, P08 suggested thresholds up to 20 stating, \textit{"For low-dimensional data, anything above 1 or 2 is too much, but for high-dimensional data...even up to 20 might be acceptable."}
These variations highlight the challenge of a universal epsilon range and the need for context-sensitive evaluations.

\paragraph{Delta:}
Experts agreed that delta should be \textbf{very small}, though specific values varied. P01 suggested \(10^{-5}\) to \(10^{-6}\) as common practice, stating, \textit{"We just want delta to be really small."} P03 and P08 emphasized its dependence on dataset size. P03 cited \(1/\sqrt{n}\) as a practical guideline, while P08 proposed \(1 / (\log n \cdot n)\). P05 favored extremely small values (e.g., \textit{".000X"}) to maintain privacy while improving accuracy.  
P01 also noted that larger databases may require even smaller deltas for meaningful privacy protection. 

\paragraph{Unit of Privacy:} Experts emphasized tailoring the unit of privacy to specific contexts. Four experts (P04, P08, P09, P10) favored a user-level focus, with P08 stating, \textit{"A normal range is anything that is individual level or a proxy for an individual."} P03 and P05 highlighted stakeholder-driven, context-dependent ranges. P09 proposed time-based units, suggesting bounds from \textit{"one day to one month or a quarter."}

\paragraph{Utility Information:} Experts emphasized that utility metrics {vary} based on data characteristics and intended use. P03 phrased utility in terms of the sizes of groups which can be reliably measured in a DP data release, with group sizes of 100 corresponding to good utility, and 50 still acceptable but 10 problematic: \textit{"If you hit 100, you're reasonable. If you hit fifty, you're good."} P08 stressed its dependency on use cases, stating, \textit{"For utility, it depends on how the data's going to be used... some may accept lower utility, while for others, high utility is critical."} P09 suggested a median relative error of $\leq 5$--$6\%$, warning against excessive hallucinated data.

\paragraph{Algorithm Hyperparameters:} Experts have varied views on normal ranges. P04 emphasized a good balance between bias and variance for noise scale, while P06 mentioned heuristic ranges for clipping norms, typically between 0.1 and 1. P09 stressed output thresholds’ role in maintaining data quality, especially for large datasets.

\paragraph{Empirical Privacy Metrics:}
P06 suggested using percentiles, minimums, maximums, and standard deviations to quantify risk, stating, "\textit{it helps to quantify the risk in terms of percentiles... giving people an average gives a clearer picture.}" They also cautioned that averages may not reflect individual experiences.

\paragraph{Privacy Interpretations or Semantics:}
For frequentist interpretations, the normal range for Type 1 errors is conventionally set at values like 0.01, 0.05, or 0.1. Power (Type 2 error) is considered meaningful between 0.4 and 0.7, and strong between 0.1 and 0.3. The range provides interpretable guarantees, since statisticians are used to thinking of uncertainty in terms of Type 1 and Type 2 error.

\subsection{Target Audiences}
\label{sec:Target audience}
To understand for whom each parameter matters, we asked experts whether they considered each parameter to be “very important,” “somewhat important,” or “not important” for the general public, and then repeated the question for technical users and data analysts. We also asked them to explain why.

These questions enabled us to capture expert opinions on how different stakeholders, both general and technical audiences, could benefit from seeing each parameter on a potential DP label. We then synthesized their responses and areas of consensus and disagreement on the appropriate target audience(s) for each parameter.
We organize our synthesis into three categories of audience relevance: (1) parameters considered relevant to both general and technical audiences; (2) parameters viewed as primarily relevant to technical users; and (3) parameters with mixed or contested relevance.
To simplify reporting, we grouped expert responses into two categories: responses of \textit{``very important''} and \textit{``somewhat important''} were combined to indicate that a parameter was considered important; responses of \textit{``not important''} were kept distinct. This approach allowed us to summarize consensus on whether a parameter was viewed as worth communicating to a particular audience.

Table~\ref{tab:dp_audience} summarizes expert recommendations on the relevance of each parameter for different audiences. Each icon reflects the level of consensus among experts, and mention counts are provided in parentheses for context.

\begin{table}[h!]
\centering
\small
\begin{tabular}{p{2.5cm}p{2.5cm}p{2.5cm}}
\toprule
\textbf{Parameter} & \textbf{General Audience} & \textbf{Technical Audience}  \\
\midrule
Epsilon &  \ding{51} (8 yes, 2 no) & \ding{51} (9 yes, 1 no)  \\
\hline
Delta & \ding{55} (1 yes, 2 no) & \ding{51} (3 yes) 
\\
\hline
Unit of Privacy & \ding{51} (6 yes, 2 no) & \ding{51} (7 yes) 
\\
\hline
Utility Information & \ding{72} (2 yes, 3 no) & \ding{51} (5 yes) 
\\
\hline
Mechanism & \ding{55} (4 no) & \ding{51} (4 yes) 
\\
\hline
Algorithm Hyperparameters & \ding{55} (4 no) & \ding{51} (4 yes) 
\\
\hline
Deployment Model & \ding{51}(3 yes) & \ding{51}(3 yes) 
\\
\hline
Empirical Metrics & \ding{51} (2 yes) & \ding{51} (2 yes) 
\\
\hline
Privacy Semantics & \ding{55} (1 no) & \ding{51} (1 yes) 
\\

\bottomrule
\end{tabular}
\caption{Expert recommendations on the relevance of each parameter for different audiences. (\ding{51} indicates "relevant"; \ding{55} indicates "not relevant"; \ding{72} indicates mixed views on relevance)}
\vspace*{-20pt}
\label{tab:dp_audience}
\end{table}

\subsubsection{Parameters  Relevant to Both Audiences}

\paragraph{Epsilon:} For the general public, 4 experts (P01, P05, P08, P12) viewed epsilon as very important, with P01 stating, "\textit{epsilon would be very important, it controls what is at the crux of the privacy guarantees.}" Another 4 experts (P04, P06, P07, P09) considered it somewhat important, and 2 experts (P03, P11) said it was not important, citing its technical complexity. For data analysts and technical audience; 7 experts (P01, P03, P04, P05, P06, P08, P12) considered epsilon very important, with P05 noting, "\textit{Epsilon is crucial for analysts as it’s their surface of interaction with DP, helping them allocate it across queries.}" Two other experts (P07, P09) viewed it as somewhat important, and one expert (P11) noted that it was not important.


\paragraph{Unit of Privacy:} This parameter received the most consensus, with most experts agreeing it matters for both audiences. For the general public; 6 experts (P04, P05, P07, P08, P09, P10) considered the unit of privacy very important, with P07 stating, "\textit{It is the one technical thing I would like the general public to understand.}" Experts (P03, P12) considered it not important, arguing it is too complex.  For data analysts and technical audience; 7 experts (P03, P04, P05, P07, P08, P10, P12) found it very important, with P07 stating that "\textit{for data analysts and technical users, what we are protecting is very important.}"

\paragraph{Deployment Model
:} Three experts (P01, P03, P05) identified deployment models as very important for both general and technical audiences. P01 emphasized its significance for public understanding of data flows and privacy risks stating that "\textit{It really changes things like who can see your data at various stages.}" The same experts also highlighted its relevance for technical audiences in assessing utility, accuracy, and threat models.

\paragraph{Empirical Privacy Metrics:}
Despite fewer mentions, those who addressed this parameter saw it as meaningful for both groups. For the general public, 2 experts (P06, P11) viewed these metrics as very important, citing that real-world attacks are more understandable than abstract guarantees. P11 noted, "\textit{Empirical attacks are easier to explain... they feel more concrete because they involve real data rather than abstract mathematical concepts.}"  For data analysts and technical audiences, P06 and P11 described empirical privacy metrics as very important, emphasizing that even technical audiences need empirical validation.

\subsubsection{Parameters Primarily Relevant to Technical Audience}
\paragraph{Delta:}
For the general public, only one expert (P01) considered delta somewhat important noting, "\textit{it’s not really driving things, but it’s sort of like important that it isn’t in the bad case.}" Two experts (P03, P05) said it was not important, with P05 stating, "\textit{If there’s agreement around what it should be and it’s set very small, then there’s no need to convey it.}" For technical audiences, 3 experts (P01, P03, P05) considered delta very important, especially for data curators.

\paragraph{Mechanism:}
For the general public; 4 experts (P01, P08, P10, P12) agreed mechanisms are not important to the general public, with P01 stating, "\textit{This is more technical content than they are able to internalize and process, I think it will add more
confusion than it clarifies.}" For data analysts and technical audiences; Some experts considered it very (P10 \& P12) or somewhat important (P01 \& P08). P12 emphasized, "\textit{The actual mechanism being used would be particularly important to them... for example, the scale term for Gaussian or Laplace noise.}"

\paragraph{Algorithm Hyperparameters:}
Four experts (P04, P06, P08, P09) considered algorithm hyperparameters very important for analysts and technical audiences but not relevant for the general public. As P04 noted, "\textit{the public doesn’t need details on noise scale or distribution,}" while P06 emphasized that such parameters "\textit{help you debug.}"

\paragraph{Privacy Semantics:}
P11 viewed privacy semantics as very important for analysts and technical audiences—especially policymakers and privacy professionals—but not important for the general public, explaining that, "\textit{There are a lot of limitations on what we can explain about privacy guarantees to the general public in a way that's useful,}" but added that, "\textit{frequentist interpretations are very important... especially for the privacy community.}"

\subsubsection{Parameters with Mixed or Contested Relevance}

\paragraph{Utility Information:}
Expert opinions on the relevance of utility information varied widely. For the general public, two experts (P03, P05) considered it very important, emphasizing that people care about accuracy. P05 stated, "\textit{People care about accuracy too.}" In contrast, three experts (P08, P09, P10) found it not important, with P10 remarking that, "\textit{Technical users will care a lot about the utility.}"  For technical users, four experts (P03, P05, P09, P10) rated utility as very important, while P08 considered it somewhat important. P05 explained, "\textit{Accuracy implications are very important for analysts... They have an interest in knowing to what extent the results of their analysis will helpfully inform that real-world purpose.}"

\subsection{Parameter-Specific Presentation Formats}
\label{sec:parameter_presentation}
We asked experts how each parameter could be more effectively presented within a DP label. Their feedback included specific suggestions for visual encoding---such as text, numbers, or range indicators---as well as preferences for placing parameters in the primary or secondary layer of a two-layer label. These expert insights were based on structured interview questions about where each parameter should appear and how it should be displayed.

To synthesize these insights, we categorized suggestions into three types of presentation formats; text explanations (e.g., plain-language descriptions), numbers (e.g., raw values like $\epsilon = 1.0$), and range indicators (e.g., color-coded or high/medium/low ranges). 
Table~\ref{tab:display_summary} summarizes expert suggestions by parameter, indicating the recommended formats, how many experts supported each, and the preferred layer of presentation.

Importantly, despite offering these concrete suggestions, many experts acknowledged the uncertainty about how best to present each parameter---underscoring the complexity of DP communication and the need for continued research to refine how each parameter is best presented. As P01 reflected \textit{``There may not
necessarily be one single unified explanation for each of these parameters, but I do think it should include some type of explanation so that users can interpret it.''} This section does not prescribe a single best format but instead highlights areas of consensus and divergence that can inform future design iterations and empirical testing. 

\paragraph{Text:}
Experts recommended text explanations for abstract parameters like unit of privacy, mechanism, deployment model, and empirical privacy metrics. P03 stated, “\textit{a simple text explanation of what aspect of the data is being protected}” is best for the unit of privacy. P01 suggested written descriptions for mechanisms stating “\textit{researchers can understand more, you can just tell them something about the algorithm.}”  For the deployment model, P01 and P03 also recommended plain-language text to describe where data resides or flows, e.g., “\textit{Does the data leave your device?}” P11 supported using text to present empirical metrics, noting that empirical privacy losses are easier to communicate in natural language than with formulas or numeric ranges.

\paragraph{Numbers:} Numeric formats—such as displaying raw values like $\epsilon = 1.0$—were primarily suggested for epsilon, delta, and privacy semantics. Three experts (P01, P05, P11) explicitly recommended numeric presentations. For epsilon, P01 and P05 suggested numeric values be shown, especially for technical users. However, P01 noted that numerical values alone may lack meaning without accompanying textual explanations. For delta, P01 indicated that a numerical value could be helpful when placed in context. P11 recommended numeric interpretations for privacy semantics to aid expert audiences, particularly for explaining frequentist interpretations and policy-relevant guarantees.

\paragraph{Indication of range:}
Color-coded or categorical range indicators (e.g., “high,” “medium,” “low”) were also recommended by experts for parameters like epsilon, delta, and utility information. Four experts (P01, P03, P06, P07) suggested that abstracting numeric values into ranges or visual symbols could help users interpret privacy strength with rangess calibrated to different models or data releases. P06 recommended a scheme “\textit{like the energy consumption ABC kind of thing}” to make privacy more intuitive. P01 proposed applying categorical color labels to delta values—“\textit{You can do colors like high, medium, low delta or something.}” For utility information, P03 inferred that users might better understand data usability if given an approximate group size, e.g., “small,” “medium,” or “large,” perhaps aided by graphic elements.

Beyond format types, a few experts shared interface-level suggestions for improving how parameters are accessed and prioritized. P05 and P09 recommended interactive display features—such as drop-downs or clickable links—to let users expand or drill down into detailed information. P04 emphasized that the \textit{order} of parameter presentation also matters, suggesting that ``\textit{unit of privacy and privacy loss parameters should appear at the top, with relaxations listed below.}''

\begin{table}[h!]
\small
\centering
\begin{tabular}{p{1.4cm}p{0.5cm}p{1cm}p{0.85cm}p{1.2cm}p{0.9cm}}
\hline
\textbf{Parameter} & \textbf{Text} & \textbf{Numbers} & \textbf{Range Indication} & \textbf{Primary Layer} & \textbf{Second Layer} \\
\hline
Epsilon & ••• & •• & •• & ••••• & • \\
\hline
Delta & • & • & • & ••• & --- \\
\hline
Unit of Privacy & ••• & --- & --- & ••••••• & --- \\
\hline
Mechanism & • & --- & --- & • & ••• \\
\hline
Deployment Model & •• & --- & --- & ••• & --- \\
\hline
Utility Info & • & --- & • & ••• & ••• \\
\hline
Empirical Metrics & • & --- & --- & •• & • \\
\hline
Privacy Semantics & • & • & --- & --- & • \\
\hline
Algorithm Hyperparameters & --- & --- & --- & --- & •••• \\
\hline
\end{tabular}
\vspace{4pt}
\caption{Expert Suggestions for Presentation Format and Layer Placement (by Parameter). One dot (•) represents one expert suggestion.}
\label{tab:display_summary}
\end{table}

\subsection{Expert Recommendations for the DP Label}
\label{expert-recommendations-Label-Design}
We sought expert recommendations on the overall design of our proposed DP label. In general, we asked open-ended questions to solicit independent suggestions for how to design a label, with one exception: \emph{we} proposed a two-layer DP label structure inspired by~\cite{emami2021informative} during the interview, to solicit expert feedback on it; experts did not independently suggest a two-layer design. Wherever possible, we clarify whether a suggestion aligned with or extended beyond our proposed label structure.

\paragraph{Standardized Contents and Formats:} 
Eleven experts praised our idea of a DP label with \textbf{standardized contents} for DP communication.
As P08 noted, \textit{"Every data release provides different parameters, having a standard would make people's lives so much easier."}
Five experts (P03, P06, P09, P11, P12) emphasized the importance of \textbf{standardized formats} aligned with existing tools and workflows. 
P09 recommended using Markdown format for broader accessibility because~\textit{"markdown is easily readable in a GitHub or GitLab repository"}.
P06 suggested integrating DP labels for machine learning data releases into existing practices, like model cards on the Hugging Face platform, to \textit{"ensure consistency and visibility"}.
P12 highlighted the importance of uniformity, suggesting the label should be \textit{"recognizable across different formats... so people know right away it’s the formal DP label."}

\paragraph{Layered Structure:}
Experts unanimously supported our proposed layered structure for DP labels.
Ten experts favored the {two-layer structure} proposed by our team: The primary layer provides high-level summaries for non-experts, while the secondary layer offers technical details for advanced DP users. 
P05 emphasized the value of this approach, stating, \textit{"The primary layer helps people understand the space of things they might need to know... researchers can click to access more details."}
Two experts (P01, P11) proposed adding \textbf{a third, intermediary layer} to bridge the gap between technical users and the general public (i.e., semi-technical audience). P11 noted the need for differentiation, explaining, \textit{"There’s a significant difference between a demographer and the general public... an intermediate category could address this."}
Two experts (P05, P09) desired an \textbf{interactive interface} to easily navigate different layers. P05 described a structure with \textit{"a high-level primary layer and the ability to drill down for more details."} 
There is an expert consensus that layered structure can balance the DP label's accessibility and depth.

\paragraph{Placement for Accessibility, Visibility, and Relevance:}
When asked about where to place the DP labels, experts offered various recommendations. 
Seven experts (P03, P06, P08, P09, P10, P11, P12) favored placing the label \textbf{alongside data releases}. 
P10 noted, \textit{"It makes sense to stick the label on the website where the dataset lives and describes the release."}
Another seven experts (P01, P03, P04, P05, P09, P10, P12) advocated \textbf{embedding the label directly into apps or services} where user interactions happen. P05 recommended placing the label in privacy centers within apps, stating that \textit{"a [DP] label could be the first summary users see"}. P04 suggested interface areas like app permissions or URL bars to increase visibility. 
Three experts (P05, P06, P12) recommended housing labels in \textbf{dedicated databases or registries} to provide centralized access. P12 noted, \textit{"Having like a secondary database too, we could look these things up."}
Furthermore, experts recommended including DP labels in \textbf{blog posts, press releases, or documentation} about specific data releases to improve transparency (P01, P07, P11), and integrating DP labels for machine learning data releases into existing formats like \textbf{model cards} (P06, P09) for relevance. 
These recommendations reflect a shared view that the placement of DP labels should maximize accessibility, visibility, and relevance.

\paragraph{Balance Between Simplicity and Education:} 
P04 believed that a simple label can help adoption, advising \textit{"Keep it simple so that it's easy to adopt."}
Experts (P01, P03, P11) also suggested that visual elements can simplify complex information, with P01 recommending \textit{"something picture-based or color-based, but not overly technical."}
Other experts (P10, P12) felt that DP labels should provide adequate DP education to assist user comprehension. 
P10 commented, \textit{"Education is key to making the label understandable"}. 
P06 wanted the DP label to prevent potential misinterpretations, suggesting \textit{"warnings... may be needed for cases where the label doesn't account for issues like data correlation."}
Overall, experts expressed different opinions on how to maximize the understandability of DP labels while keeping the label information accurate and rigorous, surfacing the need to balance between simplicity and education.

As a starting point for future work on DP label design and validation, we include our initial design of the expert-informed DP label in Appendix \hyperref[app:appendixE]{E}.

\section{Discussion}
Effective communication of differential privacy (DP) remains a significant challenge. 
Previous research focused on how to better explain obviously important parameters (e.g. epsilon ($\epsilon$)) to non-expert audiences~\cite{nanayakkara2023chances}, without considering the fact that explaining one or two DP parameters in isolation cannot adequately convey the real-world privacy protection of DP deployments.
Our study with DP experts addresses this gap by focusing on \textbf{what} parameters are essential to accurately describe the privacy guarantees of DP deployments.
It also partially tackles \textbf{how} to properly explain DP parameters through the initial design of a standardized DP label. We discuss our contributions in relation to prior work, as well as open questions in improving DP communication as follows. 

\paragraph{Towards Comprehensive DP Communication Standards:}
Unlike existing work explaining isolated parameters~\cite{nanayakkara2023chances}, our interview study with DP experts seeks to holistically communicate the privacy guarantees of DP deployments. 
As detailed in ~\ref{sec:results:parameters}, we identified a comprehensive list of parameters necessary to accurately convey the real-world privacy protection of a given DP deployment. Particularly, this list contains several critical, yet often overlooked parameters in ~\ref{sec:results:overlooked}.
We also articulated the specific considerations for each of these parameters, including their importance in determining privacy guarantees, possible normal ranges, target audiences, and  methods of presentation.
Our work provides the DP community with guidance on what should be included to describe the privacy guarantees of DP deployments, laying the foundation for future standardization efforts to improve DP communication.

\paragraph{Privacy Labels to Improve DP Communication:} Inspired by the approach of privacy labels~\cite{kelley2009nutrition,kelley2010standardizing} to standardize privacy information communication, we designed a specialized privacy label for DP by synthesizing expert feedback from this study (Appendix \hyperref[app:appendixE]{E}).
Our initial label design draws from prior work~\cite{cummings2021need,ashena2024casual} to incorporate multiple effective communication techniques (e.g. text explanations, visual elements) to improve the clarity and accessibility of DP information.
This DP label includes two layers similar to the proposed privacy label for IoT~\cite{emami2020ask}: simple summaries for the general public and detailed explanations for more technical users; 
and interactive interface to help users navigate between layers.
This two-layer design particularly addresses the needs of our target audience --- technical users who require technical details to comprehensively assess the privacy implications of a given DP deployment.

\paragraph{Bridging the Gap Between Theory and DP Deployments:}
Many existing DP explanations rely on theoretical guarantees, often abstracted in ways that make them difficult to interpret in real-world scenarios~\cite{dwork2014algorithmic,obar2020biggest}. For example, risk assessments often omit key factors such as dataset size, adversarial capabilities, or the impact of repeated queries. Practitioners find it challenging to gauge actual privacy risks.
Our study bridges this gap by emphasizing empirical privacy metrics and other contextually relevant parameters in our DP label, helping practitioners assess DP performance in practical deployments.
This enhances DP communication in operational settings, ensuring that privacy guarantees are both theoretically sound and practically meaningful.

\paragraph{Enhancing Transparency and Trust in DP Deployments:} Technical transparency is crucial for building trust and promoting DP adoption. Current projects like the Oblivious Privacy Deployments Registry~\cite{oblivious2024epsilonregistry} and Wikimedia Foundation's DP dataset documentation~\cite{wikimedia2024pageviews} primarily record privacy parameters. However, these documents often lack clear explanations about key aspects like privacy units and post-processing effects. Desfontaines' informal list~\cite{desfontaines2024realworld} adds more properties to increase transparency, but still misses some important parameters. As Cummings et al.~\cite{cummings2023centering} point out, transparency without clear communication can backfire, leading to misunderstandings and decreased trust.
We build on these works by focusing on both transparency and interpretability. Our proposed DP label records essential DP parameters and presents them in an accessible way. This approach connects theoretical DP guarantees with practical deployments and supports responsible and informed adoption of privacy-preserving technologies.

\paragraph{Open Question: Setting Parameter Ranges:}
The question of how to set $\epsilon$ is as old as DP itself~\cite{dwork2006calibrating}, and experts in our study identified similar issues with many other parameters. Appropriate settings for many parameters depend on context, making the problem even more challenging. The development of evidence-based guidance for how to set these parameters is an open research question. Approaches based on theory alone~\cite{lee2011much} have not yielded satisfying answers, and we believe that the most promising avenues would consider parameter settings in the social context of the associated deployment---for example, by taking into account utility requirements and adversary models specific to the deployment~\cite{cummings2024attaxonomy}.

\paragraph{Open Question: \textbf{How} to Communicate:}
This study focuses on \emph{what} to communicate (rather than \emph{how}). In an attempt to answer both questions at once, we sought experts' recommendations for communicating DP parameters to both technical and general audiences. Unfortunately, the major point of consensus among the experts we interviewed was: \emph{how to communicate with the general audience} remains a major open question. A significant amount of prior work has studied how to communicate key parameters to end users~\cite{bullek2017towards, franzen2022private, wen2023influence, karegar2022exploring, xiong2020towards, xiong2022using, xiong2023exploring, nanayakkara2023chances, ashena2024casual, kuhtreiber2022replication}; we hope our work will motivate similar studies for other important DP parameters.

\section{Limitations and Future Work}
\label{sec:limit-future-work}

\paragraph{Stakeholders in DP deployments.}
Our study focused on a single group of stakeholders: DP experts who may want to use a label to evaluate the technical strength of the DP guarantee associated with a deployment. Typical DP deployments include many different stakeholders with different goals and levels of expertise, including data subjects, data curators, data analysts, system engineers, compliance and legal officers, regulators, policy and decision makers, and auditing or watchdog groups.
Since each group of stakeholders represents a different population, the insights from our study likely do not generalize to all of the other groups. Further research is needed to identify and address the needs of each group, and to inform methods for ensuring transparency with each of them.

\paragraph{Study design and recruitment.}
Our qualitative study design and recruitment strategy produced rich insights about transparency in DP deployments, but the context and design of our study also has inherent limitations. Our choice of a qualitative study was motivated by the desire to enumerate important parameters for transparency in DP guarantees---since no such list previously existed, a structured, quantitative study on the relative importance of these parameters was not possible. In addition, recruiting DP experts for any type of study is challenging, since the pool of experts in this specialized field is relatively small; recruiting enough experts to obtain meaningful quantitative results is thus not feasible. Our purposive and snowball sampling may have introduced selection bias, but was necessitated by the challenges of recruiting DP experts.

\paragraph{Context of DP deployments.}
Full transparency in DP deployments requires more than communicating important parameters of the DP guarantee. For example, central-model deployments may be subject to data breaches, so real-world privacy rests on security of the stored data in addition to the parameters of the DP guarantee. The blurred line between DP parameters and other system properties affecting real-world privacy was reflected in a focus on threat modeling and deployment models in our study. A complete approach for \emph{context-inclusive} transparency in DP deployments will likely need to integrate with frameworks like like contextual integrity~\cite{nissenbaum2009privacy, nissenbaum2004privacy} and human-centric approaches for enumerating and addressing privacy threats like the NIST Privacy Risk Assessment Methodology~\cite{enterprise2020nist}.

\paragraph{Label design and validation.}
Our study sought to identify the important aspects for transparency in communicating DP guarantees, and can serve as a starting point for designing DP privacy labels for this purpose.
We plan to employ the qualitative participatory design method~\cite{spinuzzi2005methodology} with a representative sample of target users of a DP privacy label, including experts, data users, end users, and other stakeholders discussed earlier. We plan to further iterate the DP label design through large-scale evaluation studies with diverse user groups, with the goal to accommodate the communication needs of the various DP stakeholders.

\section{Conclusion}
Through an in-depth interview study with 12 DP experts, we identified the important parameters to communicate privacy guarantees of DP deployments, articulated the key considerations for each parameter, and discussed approaches to improving DP communication such as our proposed DP label .
Our findings highlight the need for standards in DP disclosures and caution against over-reliance on single parameters like $\epsilon$.
This study lays the groundwork for a standardized approach to communicating the privacy guarantees of DP deployments, fostering greater transparency and trust among DP stakeholders.

\begin{acks}
The authors would like to thank the anonymous reviewers at PETS and SOUPS for their helpful suggestions.
This material is based upon work supported by the National Science Foundation under Grant No. 2238442 and No. 2336550. Any opinions, findings and conclusions or recommendations expressed in this material are those of the author(s) and do not necessarily reflect the views of the funding agencies.
\end{acks}

\bibliographystyle{plain}
\bibliography{refs, refs_sok}

\clearpage
\appendix

\section*{Appendix}

\section*{Appendix A: Eligibility Survey}
\label{app:appendixA}
After displaying the IRB-approved consent form, the following eligibility questions were used to assess participants' qualifications for the study, collect informed consent, and gather demographic information. For questions that tested participants’ understanding of DP concepts, the correct answers are highlighted in bold.

\small
\begin{tabular}{p{8cm}}
\hline
\textbf{Eligibility Survey Questions} \\ 
\hline

\textbf{Consent/Basic Eligibility} \\

I have read and understood the information above. \newline No/Yes \\
\hline
I want to proceed to complete the eligibility survey for this research study. \newline
No/Yes \\
\hline
Are you at least 18 years old? \newline
No/Yes \\
\hline
Do you currently reside in the United States? \newline
No/Yes \\
\hline
Do you have at least five (5) years of research or practical experience in the field of differential privacy (DP)? \newline
No/Yes \\
\hline
Are you willing to participate in a 1-hour remote usability study via Microsoft Teams to help design a differential privacy nutrition label? \newline
No/Yes \\
\hline
\textbf{DP Background} \\
How many years of experience do you have working or researching in the field of Differential Privacy? \newline
(a) 0-5 (b) 5-10 (c) More than 10 \\ 
\hline
Which of the following best describes your current field of expertise? \newline
(a) Academic Research (b) Industry Practice (c) Policy Making (d) Other \\ 
\hline
What is your current job role? \newline
(a) Professor/Researcher in a University setting (b) Data Privacy Engineer (c) Policy Advisor/Consultant in Data Privacy (d) Data Scientist with a focus on Privacy Technologies (e) Other (please describe) \\

\hline
Releasing two differentially private statistics, one with $\epsilon_1$ = 0.1 and the other with $\epsilon_2$ = 0.5, results in a total privacy loss of: \newline
(a) 0.1 (b) 0.5 (c)\textbf{
0.6} (d) 0.05 (e) I don't know \\
\hline
In differential privacy, which value of the privacy parameter $\epsilon$ provides stronger privacy? \newline
(a) \textbf{
0.1} (b) 1.0 (c) I don't know \\
\hline
\textbf{Demographics} \\
\hline
What is your age? \newline
(a) 18-24 (b) 25-29 (c) 30-34 (d) 35-39 (e) 40-44 (f) 45-49 (g) 50-54 (h) 55-59 (i) I prefer not to answer \\
\hline
What is your gender? \newline
(a) Male (b) Female (c) Nonbinary(d) I prefer not to say \\
\hline
Do you hold a PhD in Computer Science with a focus on Differential Privacy or a related area? \newline Yes/No \\

\bottomrule
\end{tabular}

\section*{Appendix B: Interview Guide}
\label{app:appendixB}
The following section outlines the interview guidelines, which are organized into three focus areas: 
1) Identifying Key Differential Privacy (DP) Parameters, 
2) Assessing the key considerations including - reason for parameter inclusion, typical/normal ranges, and relevance of each DP parameter to various audiences, and 
3) Gathering recommendations for the presentation and placement of the DP  Label.

\small
\begin{tabular}{p{8cm}}
\toprule
\textbf{Interview Questions} \\
\midrule

\cellcolor{gray!25}\textbf{Focus 1: Identifying Key DP Parameters} \\


Can you list the DP parameters that you think are essential for inclusion in a DP Nutrition Label? \\
\hline 
In your opinion, is there a parameter whose importance is often underestimated or overlooked in discussions about DP? If so, can you elaborate on this? \\
\hline
\cellcolor{gray!25}\textbf{Focus 2: Assessment of key considerations} \\
\hline
Could you explain why you believe [parameter \#i] should be included in the DP Nutrition Label? \\
\hline

Could you give a potential argument against including the parameter? \\
\hline
What do you consider to be its 'normal range' in the context of differential privacy? \\
\hline
Why do you think this range is appropriate for this parameter? \\
\hline
Are there circumstances under which this 'normal range' might not be applicable or effective? If so, could you describe these situations or give examples? \\
\hline
For the general public, do you consider [parameter i] to be 'Very Important,' 'Somewhat Important,' or 'Not Important'? Why? \\
\hline

For data analysts and other technical users, do you consider [parameter \#i]  to be 'Very Important,' 'Somewhat Important,' or 'Not Important'? \\
\hline

\cellcolor{gray!25}\textbf{Focus 3: Structure, Presentation Placement and Design Recommendations} \\
\hline
We are considering a layered structure for the DP Nutrition Label, where the primary layer displays the most important information for the general public and non-technical users. Users can then click a link or follow a QR code to access a secondary layer with more detailed information for experts, researchers, and technical practitioners. What are your thoughts on this proposed design with the primary and secondary layers? \\
\hline
Considering the layered structure of the DP Nutrition Label, please indicate on which layer you think [parameter \#i] should be presented and provide your reasons.  \\
\hline
How do you think the parameter(s) should be displayed on the label? \\
\hline
Where should the DP Nutrition Labels be placed to ensure they are easily accessible and visible to users? \\
\hline
What are your thoughts on how the DP Nutrition Label should be designed, including its layout? \\
\hline
Do you have any other recommendations for the label that we haven’t discussed? \\

\bottomrule
\end{tabular}

\begin{figure}[h!]
    \centering
    \includegraphics[width=1.0\linewidth]{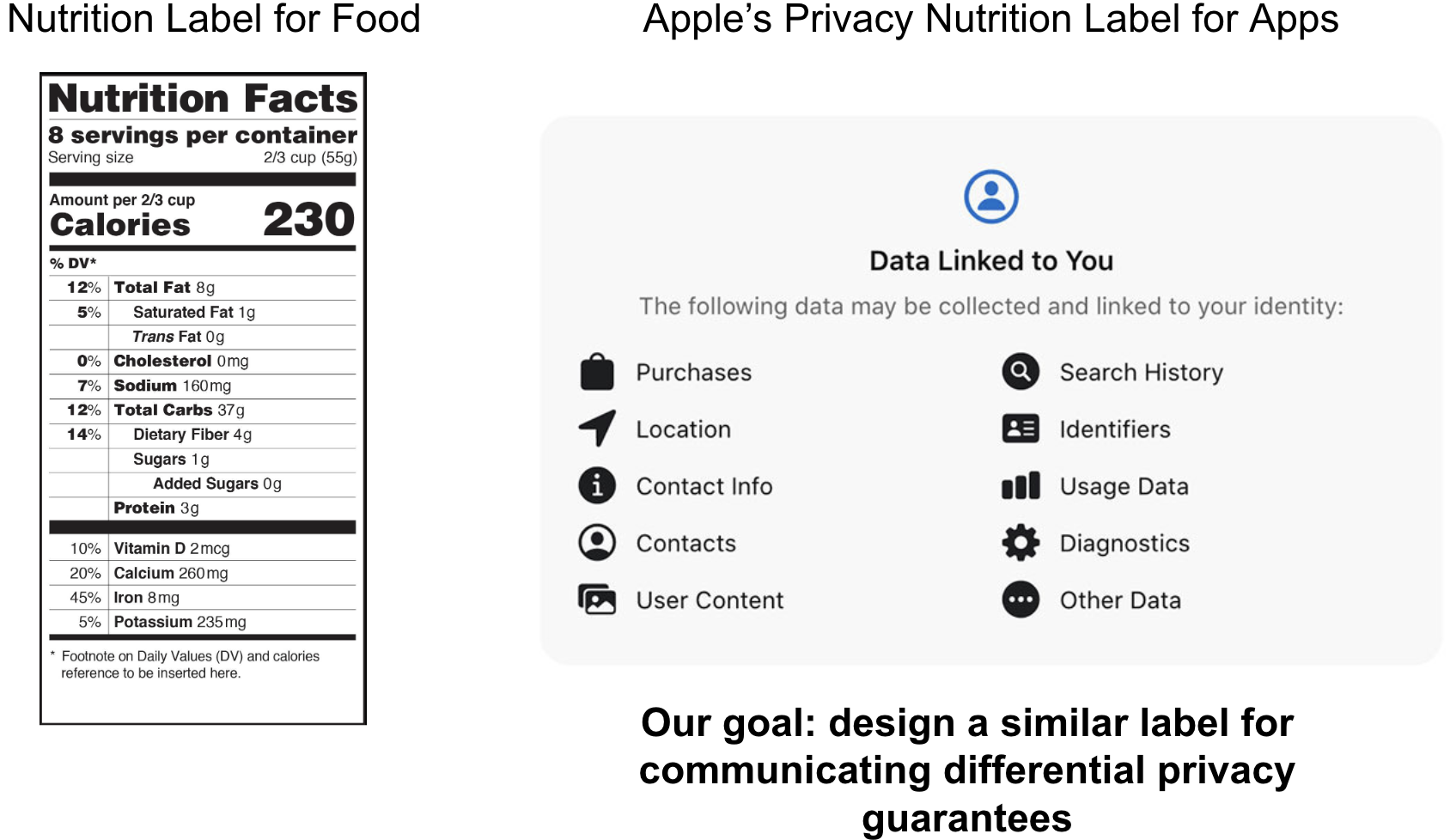}
    \caption{Example Labels}
    \label{fig:enter-label}
\end{figure}

\section*{Appendix C: More Demographics}
\label{app:appendixC}
The following tables detail the expert demographics in terms of work experience, job roles, and gender, as well as the industries and countries they represent.

\begin{table}[h!]
\centering
\begin{tabular}{ll|rlr} 
\toprule
\rowcolor{gray!35} \textbf{Gender}   & \multicolumn{3}{c}{\textbf{Role}} & \\
\midrule 
Male & 7 & & Academic Researcher & 4 \\
Female & 5 & &  Data Privacy Engineer & 4\\
\cellcolor{gray!35} \textbf{Age} & \cellcolor{gray!35} &  &  Data Scientist & 3 \\ 
\cline{1-2}
25-29 & 4 & & Software Engineer & 1 \\
35-39 & 3 & & \cellcolor{gray!35} \textbf{Country} & \cellcolor{gray!35} \\ 
\cline{4-5}
40-44 & 1 & & United States & 11 \\
55-59 & 1 & & Europe & 1 \\
\bottomrule
\end{tabular}

\caption{Expert demographics, $n = 12$}
\label{tab:expert_demographics}
\end{table}

\section*{Appendix D: List of Parameters Mentioned by Experts}
\label{app:appendixD}
Here's the list of parameters as mentioned by the experts in our study. We use "/" for the same parameters but called different names by our expert participants.


\begin{table}[H]
\begin{tabular}{p{2cm}p{2cm} p{3.5cm}}

\toprule
\textbf{Parameter Mentioned} & \textbf{Category} & \textbf{Description} \\ \\ 
\midrule
\small

Epsilon ($\epsilon$) & Privacy Parameters  & Core privacy parameter controlling the tradeoff between privacy and accuracy. \\
\hline
Delta ($\delta$) & Privacy Parameters  & Represents the probability of privacy failure in worst-case scenarios. \\
\hline
Privacy Loss Parameters (e.g., epsilon, delta) & Privacy Parameters & Additional parameters (e.g., rho for zCDP) used to quantify privacy loss. \\
\hline
Other parameters outside of epsilon, delta (e.g., rho for zCDP) & Privacy Parameters  & Alternative privacy loss metrics like rho for zCDP or Rényi DP. \\
\hline
Composition (one query or many queries) & Privacy Parameters  & Whether privacy guarantees apply to a single query or across multiple queries. \\
\hline
Query \& Mechanism used / Process for DP being applied & Mechanism Used  & Describes the method used to apply differential privacy (e.g., Laplacian or Gaussian noise). \\
\hline
Privacy Definition (e.g., Pure epsilon-DP vs zCDP) & Mechanism Used  & Specifies the type of differential privacy applied (e.g., epsilon-DP or zCDP). \\
\hline
Deployment Model & Deployment Model  & Describes how data is processed (centralized vs. decentralized), affecting privacy risks. \\
\hline
Unit of Privacy & Unit of Privacy  & Defines the level of privacy protection, such as individual or group-level privacy. \\
\hline

Budget Replenishment / Budget Reset & Privacy Parameters  & Describes whether the privacy budget is replenished or reset, impacting long-term privacy. \\
\hline
Time Frame of Guarantee (e.g., epsilon per day, per month) & Privacy Parameters  & Defines the duration of the privacy guarantee (e.g., daily, monthly). \\
\hline
Parameters that affect utility (Noise distribution, Noise scale) & Utility Information  & Defines the impact of noise and scale on the utility of the data after privacy is applied. \\
\hline

Utility Metrics & Utility Information  & Quantifies how well the data retains quality after applying DP. \\
\hline

\hline
\end{tabular}
\label{tab:parameters}
\end{table}

\begin{table}
\begin{tabular}{p{2cm}p{2cm} p{3.5cm}}

\toprule
\textbf{Parameter Mentioned} & \textbf{Category} & \textbf{Description} \\ \\ 
\midrule
\small



Worst-case vs average-case guarantee & Utility Information  & Describes whether the privacy guarantees are based on worst-case scenarios or average cases. \\

\hline
What is considered public information vs not public information & Privacy Parameters  & Defines what data is considered public versus private within the DP context. \\
\hline
Training Hyperparameters (Number of samples, etc.) & Algorithm Hyperparameters  & Describes the parameters (e.g., number of samples, iterations) affecting privacy and utility. \\
\hline
Sensitivity /Noise multiplier / Clipping norm & Algorithm Hyperparameters & Key parameters that control how noise is applied and data is clipped during processing. \\
\hline
Utility Implications / Utility and/or bias & Utility Information  & Describes how privacy parameters influence utility and potential bias in the data output. \\
\hline
Summary of Empirical Results from Privacy Attacks & Empirical Privacy Metrics & Provides results from attacks (e.g., MIA, reconstruction) assessing real-world privacy risks. \\
\hline
Non-DP Results and their impact & Empirical Privacy Metrics  & Describes the impact of non-DP data on the privacy outcomes. \\
\hline
Definition of Neighboring Datasets & Privacy Parameters  & Clarifies what constitutes a neighboring dataset for differential privacy purposes. \\
\hline
Universe of Possible Records & Privacy Parameters  & Defines the set of potential records within the DP framework. \\
\hline
Bounding User Contribution Across Releases & Privacy Parameters  & Defines how user contributions are bounded across multiple data releases. \\
\hline
Vetting of DP Implementation & Privacy Verification  & Describes the process of verifying the correctness of a DP implementation. \\
\hline
Interpretation via Frequentist or Bayesian Framework & Privacy Interpretation & Describes the statistical framework used for interpreting the privacy guarantees (frequentist vs Bayesian). \\
\hline
Privacy for Natural Groups within the Data & Group Privacy  & Ensures privacy protections apply to groups within the data, not just individuals. \\
\hline

\hline
\end{tabular}
\label{tab:parameters}
\end{table}

\begin{table}
\begin{tabular}{p{2cm}p{2cm} p{3.5cm}}

\toprule
\textbf{Parameter Mentioned} & \textbf{Category} & \textbf{Description} \\ \\ 
\midrule
\small

Where is the Data Coming From  / How is it Collected & Data Collection  & Describes the origin and collection methods for the data. \\
\hline
What Data was Published  / Who Was it Shared With & Data Sharing  & Describes the published data and who has access to it. \\
\hline
What Software was Used  / if Source Code is Available & Software and Implementation  & Describes the software used for implementing DP and whether the source code is available. \\
\hline

Publication Date  / Version Number & Data Release  & Describes when the data was released and the version number. \\
\hline
Links to Additional Materials (e.g., whitepapers)   & Additional Materials  & Provides references to related materials or research papers that support the DP deployment. \\
\hline
Goal of the Release & Data Release & Describes the intended and unintended uses of the data being released. \\

\bottomrule
\end{tabular}
\caption{Mentioned parameters}
\label{tab:parameters}
\end{table}

\clearpage
\section*{Appendix E: Prototype DP Label}
\label{app:appendixE}
\label{app:appendixE}

\begin{figure}[h]
    \centering
    \includegraphics[width=.45\textwidth]{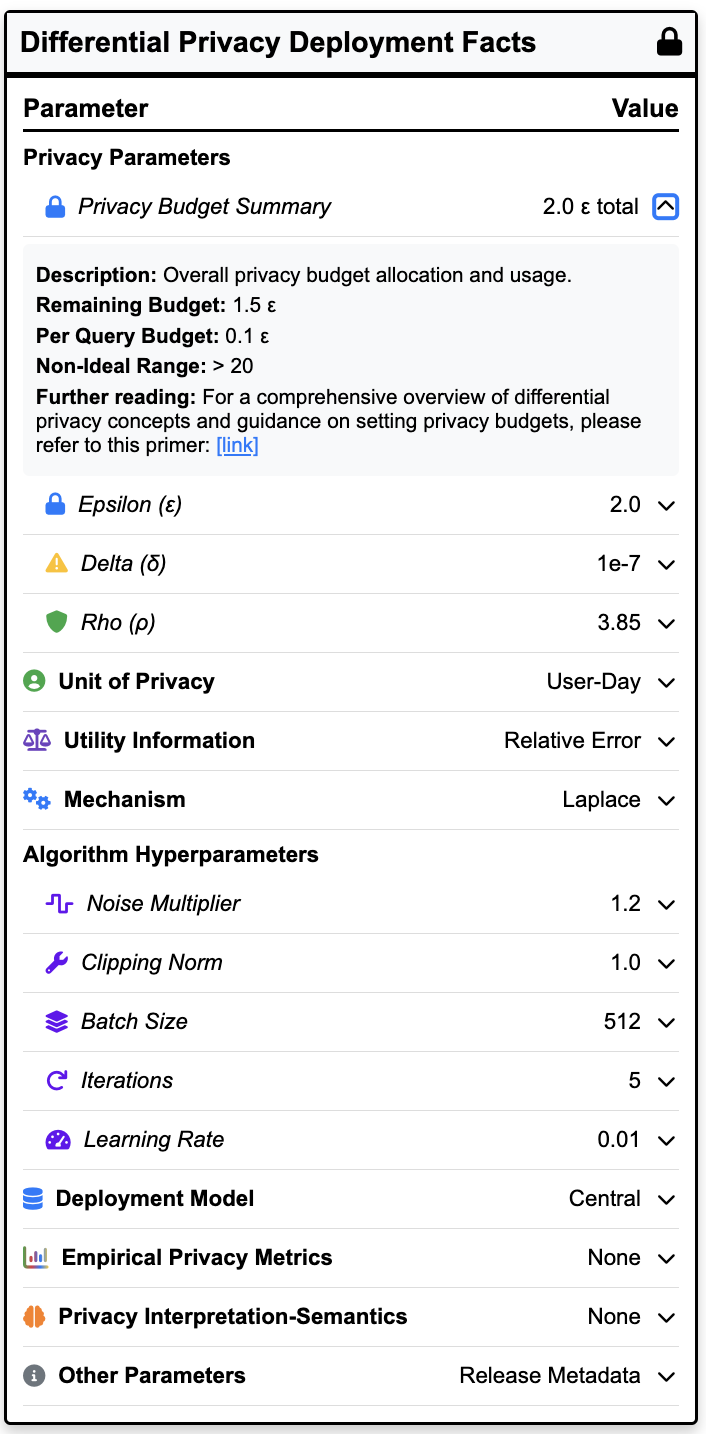}
    \caption{An exploratory visualization of expert-informed DP parameters, based on publicly documented values from Wikimedia’s deployment. This is included as a visual reference to support transparency-focused research.}
    \label{fig:appendix_dp_label}
\end{figure}

\noindent This label is presented solely as an exploratory visualization to illustrate how expert-informed DP parameters might be communicated. It reflects the consensus from expert interviews but is not evaluated or proposed as a standard. We include it to support future research.

\noindent The interactive version is available at: \\
\url{https://privacylabel4dp.github.io/Privacy-Label-for-Differential-Privacy/}

\noindent For illustrative purposes, the label uses:
\begin{itemize}
    \item \textbf{Icons} to represent parameter types (e.g., lock = $\epsilon$, gear = mechanism).
    \item \textbf{Color coding} to indicate parameter priority (e.g., green = core, blue = contextual).
\end{itemize}

\section*{Appendix F: Codebook}
\label{app:appendixF}

The codebook developed during the analysis of expert interviews includes
detailed descriptions of each theme extracted from the interview data along
with definitions.

\begin{table*}[!t]
\centering
\small
\begin{tabular}{p{0.23\textwidth} p{0.70\textwidth}}
\toprule
\textbf{Code} & \textbf{Description} \\
\midrule

\multicolumn{2}{l}{\cellcolor{gray!25}\textbf{1. Reason to Include}} \\
\midrule

Key Parameters &
Critical parameters, such as epsilon ($\epsilon$) and delta ($\delta$), that
directly impact both privacy guarantees and utility of a DP system, defining
the trade-off between data protection and usability. \\
\hline

Interpretation &
The ability to understand and compare privacy guarantees, ensuring precision
in their definition and communication. \\
\hline

Scope of Protection &
Defines the boundaries of data protection, specifying what aspects of the data
(e.g., individual records, sensitive attributes) are safeguarded under a
privacy model. \\
\hline

Transparency &
Clear communication of privacy guarantees, enabling stakeholders to verify and
trust privacy mechanisms through reproducibility, accuracy, and correctness of
implementation, ultimately building confidence in the system's privacy
assurances. \\
\hline

Utility &
Represents the usefulness of data after applying privacy mechanisms, balancing
privacy protection with the accuracy and practicality of the output. \\
\hline

Privacy Theatre &
Scenarios where certain parameters create an illusion of strong (robust)
privacy protection but fail to provide meaningful privacy or offer little
actual privacy, often leading to a false sense of trust in a system's
guarantees and neglect of other essential privacy parameters. \\
\hline

Practical Risk Assessment &
Involves evaluating risks tied to real-world harms, both positive and negative,
to ensure privacy guarantees align with practical applications and realistic
threat scenarios. \\
\hline

Understanding Utility &
Entails assessing how much usability and quality a data release retains after
applying DP mechanisms. \\
\hline

Incentive to Share &
Parameters that encourage data sharing by demonstrating that privacy mechanisms
protect sensitive information while preserving data utility. \\

\midrule
\multicolumn{2}{l}{\cellcolor{gray!25}\textbf{2. Why not Include}} \\
\midrule

None &
No valid reason exists not to disclose these parameters; they are essential
for clarity and transparency. \\
\hline

Privacy Theatre &
Parameters that mislead stakeholders by appearing to provide strong (robust)
privacy protection but fail to provide meaningful privacy or offer little
actual privacy, often leading to a false sense of trust in a system's
guarantees and neglect of other critical privacy considerations. \\
\hline

Communication Challenge &
Parameters that are too technical, complex, or lengthy to explain effectively,
making them difficult for stakeholders to understand or interpret or better
suited for detailed documentation elsewhere. \\
\hline

Doesn't Belong &
Includes parameters deemed unnecessary to disclose due to consensus,
redundancy with more expressive or critical parameters, irrelevance, or being
better suited for disclosure in other contexts. \\
\hline

Privacy Leakage &
Parameters that might inadvertently reveal sensitive or private information,
undermining privacy goals. \\
\hline

Consensus &
Parameters with no universal agreement or standardization, complicating
comparisons between mechanisms and reducing their clarity or usefulness. \\
\hline

Practical Risk Assessment &
Assesses risks tied to real-world harms but may be excluded if overly complex,
impractical, or irrelevant to stakeholders. \\
\hline

Incentive to Share &
Parameters that fail to significantly influence stakeholders' willingness to
share data or are deemed less critical. \\
\hline

Implementation &
Parameters that are challenging to implement due to technical complexity or
resource constraints or feasibility challenges. \\

\midrule

\multicolumn{2}{l}{\cellcolor{gray!25}\textbf{3. Normal Range}} \\
\midrule

Data/Problem or Algorithm Dependent &
The appropriate range for parameters varies depending on the dataset, the
problem being addressed, or the algorithm used. These ranges are inherently
context-specific and influenced by the underlying scenario. \\
\hline

Risk Tolerance &
The range reflects how much privacy risk stakeholders or data subjects are
willing to accept in real-world applications, balancing privacy protection
with acceptable exposure. \\
\hline

Specific Guidance &
Clear recommendations or constraints provided by experts, such as using very
small ranges, user-level settings, bounded/unbounded configurations, or
time-based constraints to minimize risk. \\
\hline

Consensus &
Denotes whether there is widespread agreement or standardization about the
range. While consensus may exist, discrepancies in concrete values highlight
the lack of absolute uniformity. \\
\hline

Theoretical Justification &
Ranges grounded in formal definitions or mathematical principles, providing
logical or theoretical support for their use. \\

\bottomrule
\end{tabular}
\end{table*}

\begin{table*}[!t]
\centering
\small
\begin{tabular}{p{0.23\textwidth} p{0.70\textwidth}}
\toprule
\textbf{Code} & \textbf{Description} \\







\midrule
\multicolumn{2}{l}{\cellcolor{gray!25}\textbf{4. Why the Range}} \\
\midrule

Theoretical Justification &
The range is supported by established theoretical frameworks or formal
definitions, ensuring mathematical soundness. \\
\hline

Consensus &
A range commonly accepted or standardized within the field, facilitating
comparisons and usability across implementations. \\
\hline

Data/Problem or Algorithm Dependent &
The range is determined by the specific context, such as the dataset's
characteristics, the problem's requirements, or the algorithm being employed. \\
\hline

Utility &
The range ensures a balance between maintaining privacy and retaining practical
usability of the data, optimizing both goals. \\
\hline

Stakeholder Feedback or Expectation &
Ranges are influenced by stakeholder inputs, reflecting their expectations for
balancing privacy guarantees with practical application needs. \\
\hline

Quantifies Risk &
The range allows stakeholders to assess and quantify the potential privacy
risks associated with specific parameter choices. \\

\midrule
\multicolumn{2}{l}{\cellcolor{gray!25}\textbf{5. Circumstances the normal range is not applicable}} \\
\midrule

Composition / Complexity &
In scenarios involving complex models or multiple composed queries, the normal
range may not apply due to added layers of difficulty in maintaining
consistency. \\
\hline

Utility &
If adhering to the normal range significantly reduces data usability, the
range becomes impractical for the intended purpose. \\
\hline

None &
There is no valid reason not to apply the range. \\
\hline

Data/Problem or Algorithm Dependent &
Variability in data, problem scope, or algorithm characteristics may render
the normal range irrelevant or inapplicable. \\
\hline

Implementation &
Challenges in implementing the range due to technical complexity or
limitations or resource constraints may prevent its applicability. \\
\hline

Stakeholder Feedback or Expectation &
Deviations from the normal range may be necessary if stakeholders require
different balances of privacy and utility. \\
\hline

Threat model &
The range may not apply if the threat model changes or introduces additional
risks that require a different parameter setting. \\
\hline

Consensus &
The lack of agreement or standardization in the field can make the normal
range irrelevant or inapplicable. \\

\bottomrule
\end{tabular}
\end{table*}

\begin{table*}[!t]
\centering
\small
\begin{tabular}{p{0.23\textwidth} p{0.70\textwidth}}

\multicolumn{2}{l}{\cellcolor{gray!25}\textbf{6. Important for General Public}} \\
\midrule

Very Important &
Indicates that the parameter is crucial for the general public to understand,
as it significantly impacts their trust, informed decisions (data-sharing
decisions) about data privacy and its implications, or perception of privacy
guarantees. \\
\hline

Somewhat Important &
Denotes that the parameter is moderately relevant to the general public,
providing useful context but not essential for their understanding or
decision-making processes. \\
\hline

Not Important &
Denotes that the parameter holds little to no relevance to the general public
and does not contribute meaningfully to their understanding or data-sharing
decisions. \\

\midrule
\multicolumn{2}{l}{\cellcolor{gray!25}\textbf{7. Important for Technical Users}} \\
\midrule

Very Important &
Indicates that the parameter is essential for technical users to implement,
evaluate, or optimize DP systems effectively. \\
\hline

Somewhat Important &
Suggests that the parameter is moderately relevant for technical users, aiding
their tasks but not being critical for successful implementation or
evaluation. \\
\hline

Not Important &
Implies that the parameter is not relevant or necessary for technical users in
their roles or tasks. \\

\midrule
\multicolumn{2}{l}{\cellcolor{gray!25}\textbf{8. Thoughts about layered structure}} \\
\midrule

Support &
Reflects agreement or approval that a layered structure is effective for
presenting parameters to audiences with different expertise levels, ensuring
accessibility for all stakeholders. \\
\hline

Add Third Layer &
Suggests enhancing the layered structure by introducing a third layer to tailor
the presentation of parameters for different user groups, such as adding a
deeper technical level or simplifying for non-technical users. \\

\midrule
\multicolumn{2}{l}{\cellcolor{gray!25}\textbf{9. Layer to Present Parameters}} \\
\midrule

Primary &
The top layer designed to present the most critical parameters, offering
high-level, accessible information for a broad audience. \\
\hline

Secondary &
A deeper layer intended for more detailed or technical explanations, tailored
for users with advanced knowledge (technical users) or specific needs. \\

\bottomrule
\end{tabular}
\end{table*}

\begin{table*}[!t]
\centering
\small
\begin{tabular}{p{0.23\textwidth} p{0.70\textwidth}}
\toprule
\textbf{Code} & \textbf{Description} \\
\midrule

\multicolumn{2}{l}{\cellcolor{gray!25}\textbf{10. How to Present Parameters}} \\
\midrule

Full Technical Info &
Present the complete and detailed information about the parameter, including
numerical values, descriptions, and links to supporting documentation or
research papers for technical users. \\
\hline

Color-coded &
The use of a color scheme to visually differentiate between parameter
categories, levels of risk, or importance for quick comprehension (e.g., green
for safe, yellow for caution, red for high-risk). \\
\hline

High-Level Summary &
A brief, non-technical explanation of parameters, designed to be understandable
by non-expert audiences while conveying the core concepts. \\

\midrule
\multicolumn{2}{l}{\cellcolor{gray!25}\textbf{11. Where to Place the Label}} \\
\midrule

With the Data Release &
Attaching the DP label directly to the dataset being shared to ensure immediate
access by users evaluating the data. \\
\hline

In the App &
Embedding the DP label within the application or tool that interacts with the
data, ensuring it's accessible during use. \\
\hline

In a Registry &
Storing the DP label in a centralized registry where datasets and their privacy
parameters are documented for easy reference. \\
\hline

In a Blog Post/Press Release/Docs &
Including the DP label in public communications or documentation to provide
transparency and educate broader audiences about privacy protections. \\
\hline

In a Model Card &
Integrating the DP label into a structured document (model card) that outlines
key details about the data, its use, and privacy guarantees. \\

\midrule
\multicolumn{2}{l}{\cellcolor{gray!25}\textbf{12. Thoughts on design and layout}} \\
\midrule

Color-coding &
The use of a color scheme to visually differentiate between parameter
categories, levels of risk, or importance for quick comprehension (e.g., green
for safe, yellow for caution, red for high-risk). \\
\hline

Consistency and Standardization &
The DP label follows a uniform design and structure to improve usability and
avoid confusion across datasets or tools. \\
\hline

Interactivity &
The incorporation of interactive elements, such as tooltips or expandable
sections, to allow users to explore detailed parameter information as needed. \\
\hline

Use of Visuals &
Addition of visual elements like icons, charts, or diagrams to enhance the DP
label's clarity and make complex information easier to understand. \\
\hline

Simple File Format &
The DP label should be in a format that is easy to integrate and share, such
as README. \\
\hline

Order Matters &
Structuring the DP label to prioritize the most critical parameters ensuring
important parameters are noticed first. \\

\midrule
\multicolumn{2}{l}{\cellcolor{gray!25}\textbf{13. Recommendations}} \\
\midrule

Educational Resources &
These are supplementary materials, such as tutorials or guides, to help users
understand the parameters and their implications. \\
\hline

Mirror Existing Labels &
Adopting design elements or structures from established labels (e.g.,
nutrition labels) to leverage familiarity and reduce the learning curve. \\
\hline

Checklist &
A suggestion of having a checklist format, allowing users to verify that all
critical privacy elements are addressed. \\
\hline

Seek Feedback &
Engage with users, stakeholders, or experts to refine the label's design and
content based on their needs and suggestions. \\
\hline

Simplicity &
The DP label should be straightforward, avoiding unnecessary complexity to
ensure it is accessible and user-friendly. \\
\hline

Highlight Potentially Misleading Aspects &
Clearly identifying areas where the label might unintentionally cause
misinterpretations in privacy guarantees. \\

\bottomrule
\end{tabular}
\end{table*}

\clearpage

\end{document}